\documentclass[apj, letterpaper]{emulateapj}  

\pdfoutput=1

\usepackage{amsmath}

\newcommand{\m}{\rm \,m}

\newcommand{\J}{\rm \,J}
\newcommand{\kg}{\rm \,kg}

\newcommand{\W}{\rm \, W}

\def\tt{}

\begin{document}

\slugcomment{\bf}
\slugcomment{Astrophysical Journal, accepted}

\title{Atmospheric circulation of hot Jupiters: insensitivity to initial conditions}



\author{ Beibei Liu\altaffilmark{1}, Adam P.\ Showman\altaffilmark{2}}
\altaffiltext{1}{Kavli Institute for Astronomy \& Astrophysics, and Department of Astronomy, Peking University, Beijing, 100871, P.R. China}
\altaffiltext{2}{Department of Planetary Sciences and Lunar and Planetary
Laboratory, The University of Arizona, 1629 University Blvd., Tucson, AZ 85721 USA; showman@lpl.arizona.edu}

\begin{abstract}
\label{abstract}

The ongoing characterization of hot Jupiters has motivated a variety
of circulation models of their atmospheres.  Such models must be
integrated starting from an assumed initial state, which is typically
taken to be a wind-free, rest state.  Here, we investigate the
sensitivity of hot-Jupiter atmospheric circulation models to initial
conditions.  We consider two classes of models---shallow-water models,
which have proven successful at illuminating the dynamical mechanisms
at play on these planets, and full three-dimensional models similar to
those being explored in the literature.  Models are initialized with
zonal jets, and we explore a variety of different initial jet
profiles.  We demonstrate that, in both classes of models, the final,
equilibrated state is independent of initial condition---as long as
frictional drag near the bottom of the domain and/or interaction with a
specified planetary interior are included so that the atmosphere can adjust
angular momentum over time relative to the interior.  When such
mechanisms are included, otherwise identical models initialized with
vastly different initial conditions all converge to the same
statistical steady state.  In some cases, the models exhibit modest
time variability; this variability results in random fluctuations
about the statistical steady state, but we emphasize that, even in
these cases, the statistical steady state itself does not depend on
initial conditions.  Although the outcome of hot-Jupiter circulation
models depend on details of the radiative forcing and frictional drag,
aspects of which remain uncertain, we conclude that the specification
of initial conditions is not a source of uncertainty, at least over
the parameter range explored in most current models.

\end{abstract}



\section{Introduction}
\label{Introduction}
Since the discovery of the first exoplanet around a main-sequence
star, 51~Pegasi~b \citep{mayor-queloz-1995}, almost 800 planets
orbiting other stars have been detected.  Nearly $20\%$ of them are
hot Jupiters \citep{wright-etal-2011}, giant planets orbiting within
$\sim$$0.1\,$AU of their central stars.  The physical regime of hot
Jupiters differs substantially from those of solar-system giant
planets; the presumed tidal locking leads to modest rotation rates
and permanent day and nightsides, with incident stellar fluxes
$\sim$$10^3$--$10^6$ times stronger than that received by giant
planets in our solar system.  These differences have motivated a 
flourishing research program focused on elucidating the atmospheric
dynamics of these worlds.  Current observations place meaningful
constraints on planetary radii, atmospheric composition, albedo, and
the three-dimensional temperature structure, including dayside
temperature profiles and variation of the temperature between day and
night \citep[e.g.,][]{knutson-etal-2007b,charbonneau-etal-2008,
knutson-etal-2008, cowan-etal-2007, harrington-etal-2006, 
harrington-etal-2007}.

These observations have motivated a variety of three-dimensional (3D)
circulation models of hot Jupiters \citep{showman-guillot-2002,
  cooper-showman-2005, cooper-showman-2006, showman-etal-2008a,
  showman-etal-2009, menou-rauscher-2009, rauscher-menou-2010,
  rauscher-menou-2012, dobbs-dixon-lin-2008, dobbs-dixon-etal-2010,
  lewis-etal-2010, perna-etal-2010, perna-etal-2012,
  thrastarson-cho-2010, thrastarson-cho-2011, heng-etal-2011,
  heng-etal-2011b, showman-etal-2012}.  These models obtain a
circulation pattern generally comprising several broad atmospheric
jets, including in most cases a strong eastward equatorial jet
(superrotation) with speeds up to a few $\rm km\,s^{-1}$ and westward
zonal-mean flow at high latitude.  An analytical explanation was
provided by \cite{showman-polvani-2011}, who showed that the
superrotation results from the interaction of planetary-scale Rossby
and Kelvin waves---themselves a response to the day-night thermal
forcing---with the mean flow.  When the radiative and advection time
scales are comparable, the hottest region can be displaced eastward
from the substellar point by tens of degrees longitude
\citep[e.g.,][]{showman-guillot-2002}, as subsequently observed on HD
189733b \citep{knutson-etal-2007b}.  Nevertheless, it is notable that
some 3D models produce qualitatively different circulation patterns
exhibiting a range of flow behavior \citep{thrastarson-cho-2010}.
 
An important issue in modeling atmospheric circulation is the choice
of initial conditions.  Most of the models published to date integrate
the equations starting from a rest state containing no winds, although
\citet{cooper-showman-2005} also performed an integration starting
from an initial condition containing a broad westward jet and found
that the initial condition did not strongly affect the outcome.
Nevertheless, \citet{thrastarson-cho-2010} recently performed a
detailed investigation and found that, in their model, the initial
conditions severely affect the final state.  They explored initial
conditions containing a broad eastward jet, broad westward jet, and
three-jet pattern, all with speeds of 0.5--$1\rm\,km\,s^{-1}$, and
compared this to models integrated from rest.  These
models---differing only in the initial condition---led to a wide range
of final states whose pattern of flow streamlines and the horizontal
temperature distribution (including the longitudinal offsets of any
hot or cold spots) differed significantly.
\citet{thrastarson-cho-2010} found that such extreme sensitivity to
initial condition occurred even in models whose initial conditions
differed only slightly.


From a mathematical perspective, the initial conditions of course
comprise an essential aspect of defining the overall mathematical
problem, and in many dynamical systems, the initial conditions indeed
play a crucial role in affecting the solution.  However, atmospheres
are forced-dissipative systems, and such forcing and dissipation tend
to drive the atmospheric circulation into a statistical steady state
retaining little if any memory of its initial condition.  For this
reason, many terrestrial general circulation model (GCMs)
investigating the statistical steady state of Earth's global
atmospheric circulation are integrated from rest,\footnote{This of
  course is not true for weather prediction models, which consist of
  very-short-term integrations of typically a few days, determining
  how some given observed circulation evolves with time.} with no
expectation that this choice adversely influences the results. 

A possible exception to this expected lack of sensitivity would be if
the atmospheric circulation exhibits multiple, stable equilibria
corresponding to radically different circulation patterns for an
identical set of forcing and boundary conditions.  In this case,
determining which of these multiple equilibria the atmosphere resides
in requires knowledge of its history, potentially including its
initial condition.  An example in the context of terrestrial-planet
climate is the existence of stable equilibria corresponding to
primarily ice-free and globally glaciated (``Snowball Earth'') states
\citep{budyko-1969}.  In the range of incident stellar fluxes where
both equilibria exist, the actual state occupied by the climate
depends on history \citep[for specific examples of how this works,
  see][]{pierrehumbert-2010}.  In just such a way, the question of
initial-condition sensitivity for the atmospheric circulation is
therefore perhaps best thought of as a question of whether the
atmosphere exhibits multiple rather than only one stable equilibrium.
To date, no claims have been made in the literature that the
atmospheric circulation of hot Jupiters exhibit multiple, stable
equilibria for a given set of forcing conditions, and so a sensitivity
to initial conditions \citep{thrastarson-cho-2010} is not expected {\it
  a priori}.  Nevertheless, the issue is worthy of further study.

Here, we present the results of a thorough exploration of the initial
condition sensitivity in atmospheric circulation models of hot Jupiters.
We investigate two classes of models---idealized shallow-water models
in Section~\ref{2D model}, as these have recently been used to identify
dynamical mechanisms operating in the atmospheres of hot Jupiters 
\citep{showman-polvani-2011}---and fully 3D models in Section~\ref{3D model}.
Section~\ref{conclusion} concludes.

\section{shallow water model}
\label{2D model}
\subsection{Model description}


Simplified models play an important role in understanding atmospheric
dynamical processes.  The shallow-water model, in particular, is an
idealized fluid system that has proven successful in illuminating
many aspects of the large-scale dynamics, both for Earth
\citep[e.g.,][]{polvani-etal-1995} and giant planets
\citep{dowling-ingersoll-1989, cho-polvani-1996a, showman-2007,
scott-polvani-2007, scott-polvani-2008}.  Recently, 
\citet{showman-polvani-2011} showed that, when day-night thermal
forcing is included as appropriate to synchronously rotating hot Jupiters,
the model produces a circulation with many similarities to those
emerging in full 3D models of hot Jupiters.  Here, we explore
the sensitivity of this model to initial conditions. 

We adopt a two-layer model, with constant densities in each layer;
the upper layer, of lesser density, represents the meteorologically
active atmosphere, while the lower layer, of greater density, represents
the deep interior.  In the limit where the lower layer is quiescent
and infinitely deep, this system reduces to the shallow-water equations
for the flow in the upper layer:
\begin{equation}
{d{\bf v}\over dt}+g\nabla h + f{\bf k}\times {\bf v} = {\bf R} 
- {{\bf v}\over{\tau_{\rm drag}}}
\label{momentum0}
\end{equation}
\begin{equation}
{\partial h\over\partial t} + \nabla\cdot ({\bf v}h) = 
{h_{\rm eq}(\lambda,\phi) - h\over\tau_{\rm rad}}
\equiv Q
\label{continuity0}
\end{equation}
where ${\bf v}(\lambda,\phi,t)$ is horizontal velocity, 
$h(\lambda,\phi,t)$ is the upper layer thickness, 
$t$ is the time, $g$ is the (reduced) gravity, 
$f=2\Omega\sin\phi$ is the Coriolis parameter, ${\bf k}$ is the
upward unit vector, $\Omega$ is planetary rotation rate,
$d/dt=\partial/\partial t + {\bf v}\cdot\nabla$ is the material
derivative (including curvature terms in spherical geometry), 
and $\lambda$ and $\phi$ are the longitude and latitude,
respectively.

Radiative heating and cooling are treated using a Newtonian cooling scheme,
which relaxes the upper layer thickness toward a specified
radiative-equilibrium thickness, $h_{\rm eq}(\lambda,\phi)$ over 
a specified radiative timescale, $\tau_{\rm rad}$.  To represent the
day-night heating pattern on a synchronously rotating hot Jupiter,
$h_{\rm eq}$ is chosen to be thick on the dayside and thin on the nightside: 
\begin{equation}
h_{\rm eq}=H + \Delta h_{\rm eq}\cos\lambda\cos\phi
\label{heq}
\end{equation}
where $H$ is a constant mean thickness and $\Delta h_{\rm eq}$ is the day-night
contrast in radiative equilibrium thickness. The substellar point is
at longitude $0^{\circ}$ and latitude $0^{\circ}$.  The equations
include frictional drag, represented as a linear damping of winds with
a specified drag timescale, which could represent the potential
effects of magnetohydrodynamic friction \citep{perna-etal-2010},
vertical turbulent mixing \citep{li-goodman-2010}, or momentum
transport by breaking gravity waves \citep{watkins-cho-2010}.  The
term $\mathbf{R}$ in Eq.~(\ref{momentum0}) represents momentum
transport between the layers and is
\begin{equation}
{\bf R}(\lambda,\phi,t)=
\begin{cases}
   -{Q{\bf v}\over h},  &Q>0;\\
   0, &Q<0
\end{cases}
\label{R}
\end{equation}
Air moving into the upper layer $(Q>0$) affects the upper layer's
specific angular momentum, but air moving out of the upper layer does
not.  See \citet{showman-polvani-2011} for further discussion
and interpretation of the equations.
Parameters were chosen to be representative of hot Jupiters, including
$g=20\rm\,m\,sec^{-2}$, $H=200\rm\,km$,
$\Omega=3.2\times10^{-5}\rm\,sec^{-1}$ and $a=8.2\times10^7\rm\,m$,
similar to the values on HD 189733b.  Note that the specific choices
of these parameters are not essential to the result; similar results 
would obtain were other parameter values used instead.

\placefigure{}
\begin{figure} 
  \centerline{\plotone{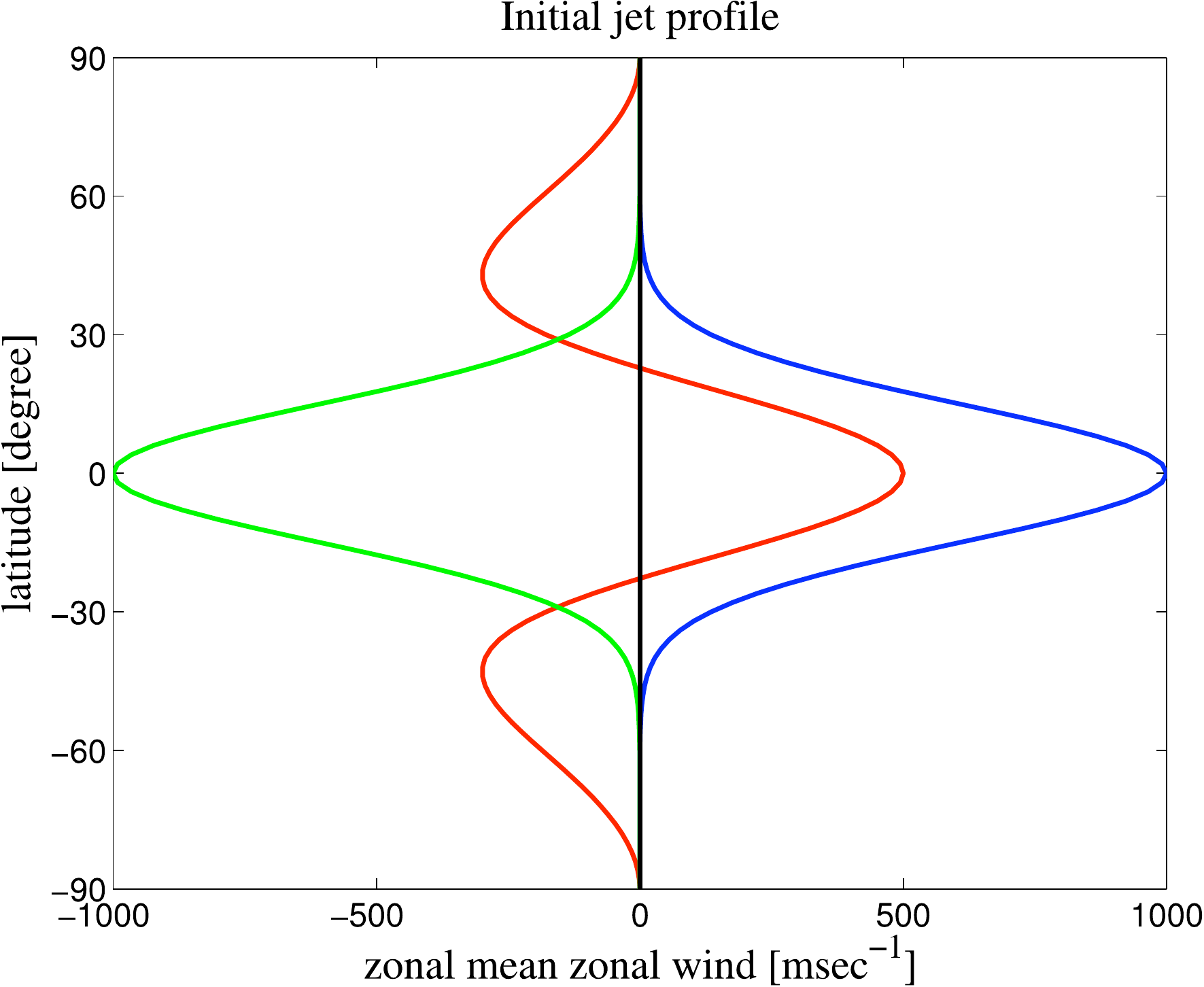}} 
  \caption{Initial jet profiles adopted in our shallow-water models,
    corresponding to a rest state (black line), eastward jet (blue
    line), westward jet (green line) and three-jet pattern (red line).}
  \label{sw-set}
\end{figure}

Our initial conditions are motivated by those of \citet{thrastarson-cho-2010}
and are shown in Figure~\ref{sw-set}.  The model is initialized with
a zonally symmetric (i.e., longitude-independent) zonal flow given by
\begin{equation}
u_{\rm initial}(\phi)=U\exp\left[-{\phi^2 \over 2\sigma^2}\right].
\label{initial-jet}
\end{equation}
where $\sigma$ is the jet half-width and $U$ is the jet speed.  We
present models whose initial zonal winds comprise eastward equatorial
jets ($U=1\rm\,km\,s^{-1}$ and $\sigma=\pi/12$, blue curve in
Figure~\ref{sw-set}), westward equatorial jets ($U=-1\rm\,km\,s^{-1}$
and $\sigma=\pi/12$, green curve in Figure~\ref{sw-set}), and a
three-jet profile (red curve), as well as models integrated from rest
(black curve).  The initial meridional velocity is set to zero and the
height field is specified to be in gradient-wind balance with the
zonal-wind field.  Note that this is only a subset of the models
explored; other initial conditions were tried as well and yielded the
same results.

\begin{table*}[!ht]
\centering
\caption{The properties of some shallow-water runs}
\begin{tabular}{|c|c|c|c|c|c|}
\hline
\hline
Name & Initial condition &$\rm \Delta h_{eq}/H$ & $\rm \tau_{rad}$ ($\rm days$) & $\rm \tau_{drag}$  ($\rm days$)  & Time ($\rm days$)\\
\hline
SW1 & Rest state & $0.01$   &     $1$    &   $1$  & 100 \\
SW2 & Eastward jet & $0.01$   &     $1$    &   $1$ & 100  \\
SW3 & Westward jet & $0.01$   &     $1$    &   $1$  & 100\\
SW4 & Combined jet & $0.01$   &     $1$    &   $1$  & 100\\
SW5 & Rest state  & $0.5$     &    $0.1$    &   $\infty$  & 1000\\
SW6 & Eastward jet & $0.5$     &    $0.1$    &   $\infty$ & 1000 \\
SW7 & Westward jet & $0.5$     &    $0.1$    &   $\infty$ & 1000 \\
SW8 & Combined jet & $0.5$     &    $0.1$    &   $\infty$  & 1000\\
\hline
\end{tabular} 
\label{tab1}
\end {table*}

We integrate Eqs.~(\ref{momentum0})--(\ref{continuity0}) using the Spectral
Transform Shallow Water Model (STSWM) \citep{hack-jakob-1992}, which
solves the equations using pseudospectral methods in vorticity-divergence
form.   We adopt a spectral truncation of T170, corresponding to a 
resolution of $0.7^{\circ}$ in longitude and latitude (i.e.,
a global grid of $512\times256$ in longitude and latitude).  The code
adopts the leapfrog time stepping method, using an Asselin filter to
suppress the computational mode.  A $\nabla^6$ hyperviscosity is applied
to each of the dynamical variables to maintain numerical stability.
All models are equilibrated to a statistical steady state.

\subsection{Results}

We explored a wide range of models with differing values of
$\tau_{\rm rad}$, $\tau_{\rm drag}$, $\Delta h_{\rm eq}/H$, and
initial condition.  A small subset of these models, which are 
illustrated in subsequent figures, are shown in Table~\ref{tab1}.

\begin{figure}[!h]
\centering
\includegraphics[scale=0.8, angle=0]{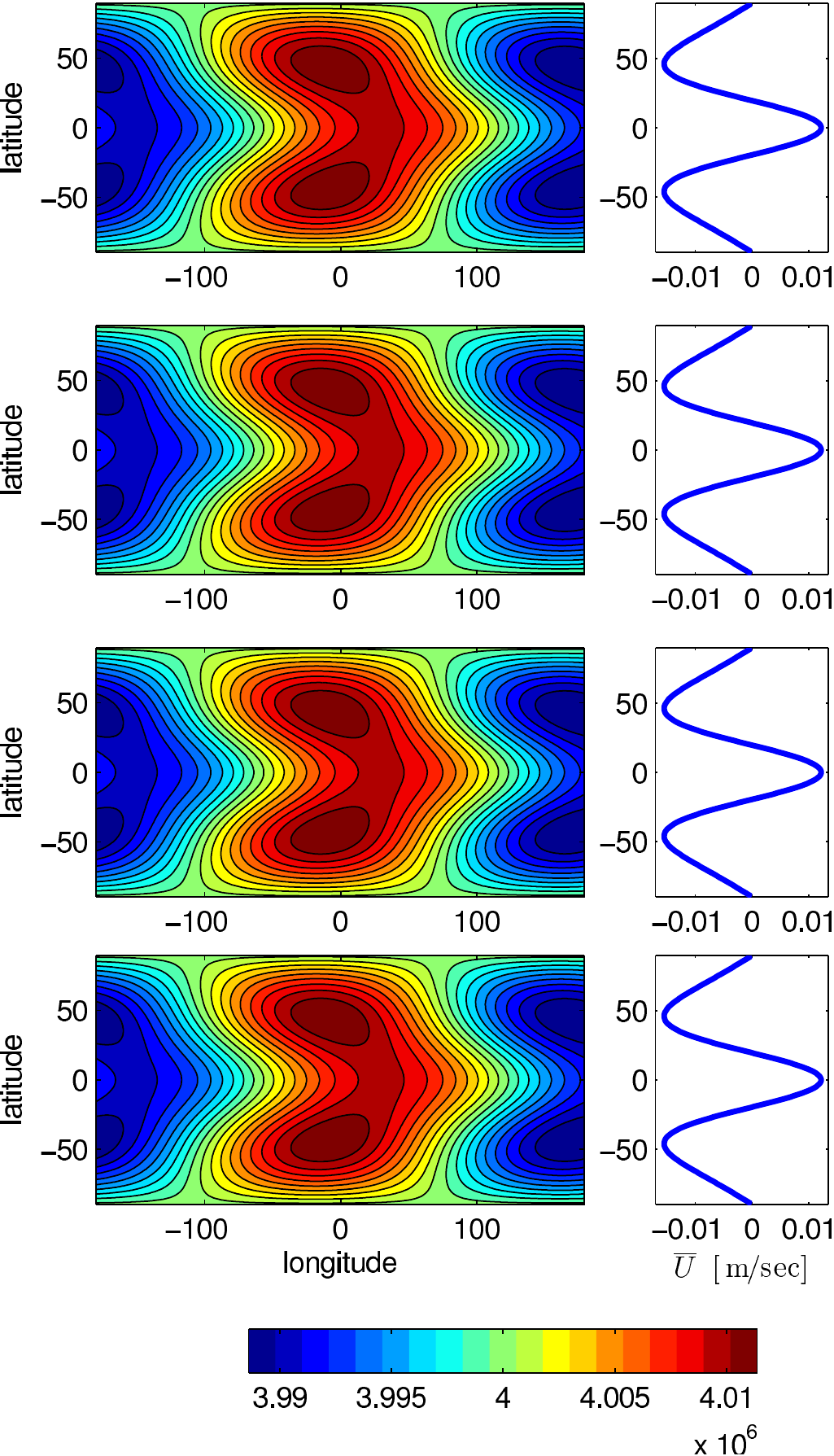} 
\caption{Geopotential ($gh$) over the globe (left) and zonal-mean
  zonal wind (right) in four low-amplitude shallow-water models
  initialized with differing initial conditions corresponding to the
  four initial conditions in Figure~\ref{sw-set}: a rest state (top
  row), eastward equatorial jet (second row), westward equatorial jet
  (third row), and three-jet pattern (fourth row).  Units of colorbar
  are $\rm m^2\,s^{-2}$.  These are models SW1, SW2, SW3, and SW4 from
  top to bottom, respectively.  These are snapshots, shown at 100
  days, after the models have reached equilibrium.  Despite the
  different initial conditions, the models all converge to the
  identical steady state.}
\label{sw-low}
\end{figure}

\begin{figure}[!h]
\centering
\includegraphics[scale=0.8]{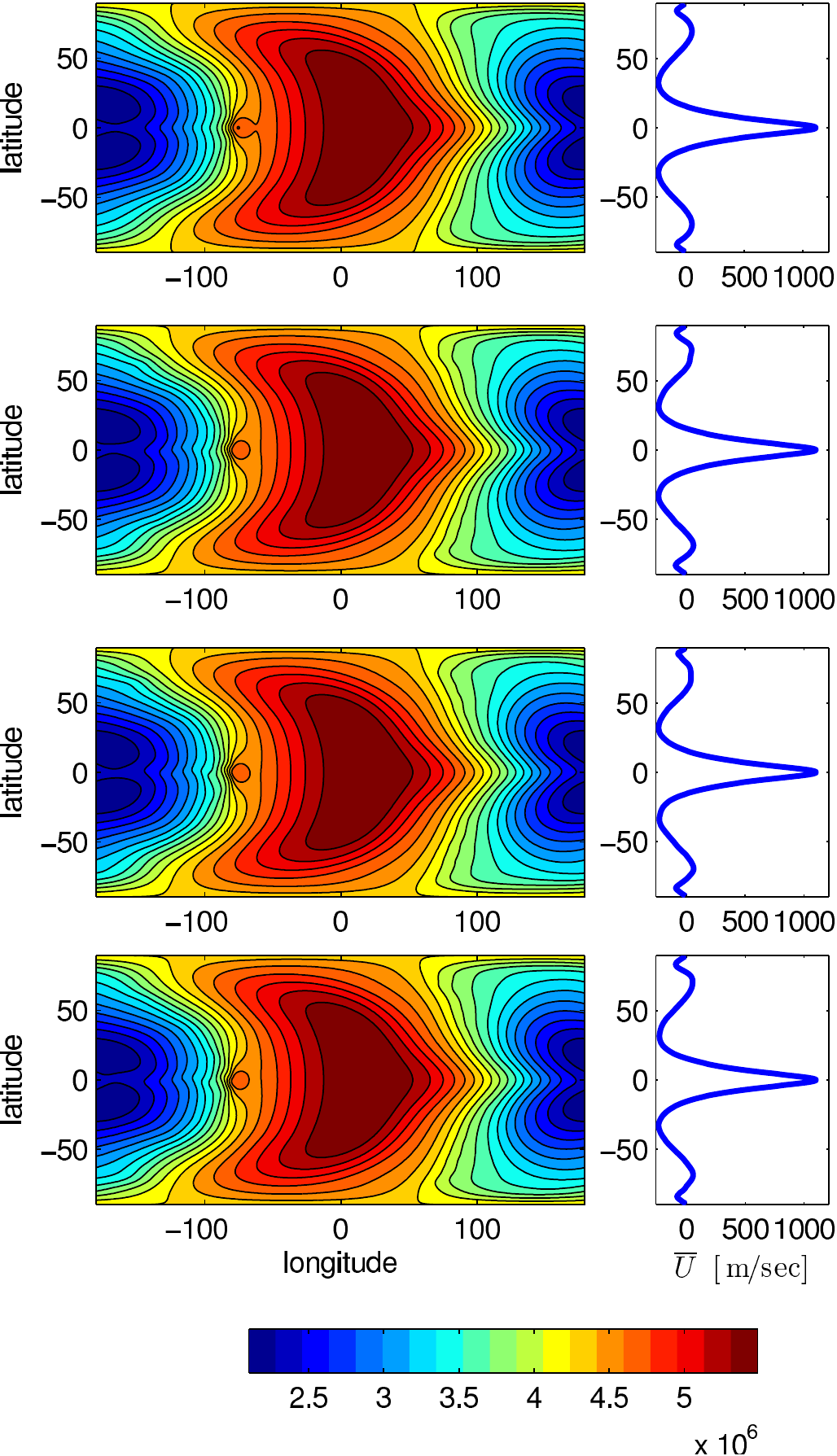} 
\caption{Geopotential ($gh$) over the globe (left) and zonal-mean
  zonal wind (right) in four high-amplitude shallow-water models
  initialized with differing initial conditions corresponding to the
  four initial conditions in Figure~\ref{sw-set}: a rest state (top
  row), eastward equatorial jet (second row), westward equatorial jet
  (third row), and three-jet pattern (fourth row).  Units of colorbar
  are $\rm m^2\,s^{-2}$.  These are models SW5, SW6, SW7, and SW8 from
  top to bottom, respectively.  These are snapshots, shown at 1000
  days, after the models have reached equilibrium.  Despite the
  different initial conditions, the models all converge to the same
  statistical steady state.  }
\label{sw-high}
\end{figure}

We find, over a wide range of parameters, that the final, equilibrated
state is independent of initial condition.  This is illustrated in
Figures~\ref{sw-low} and \ref{sw-high}.  At sufficiently low
amplitude---that is, at sufficiently low values of $\Delta h_{\rm
  eq}/H$---the equilibrated solutions are temporally steady, whereas
time variability sets in beyond a critical amplitude
\citep{showman-polvani-2010, showman-polvani-2011}.
Figure~\ref{sw-low} shows the steady-state geopotential, $gh$, and
zonal-mean zonal winds at an integration time of 100 days\footnote{In
  this paper, 1 day is defined as 86400 sec.}  in low-amplitude models
with $\Delta h_{\rm eq}/H=0.01$, $\tau_{\rm rad}=1\rm\,day$, and
$\tau_{\rm drag}=1\rm\,day$.  The four models in Figure~\ref{sw-low}
are integrated from the four initial conditions shown in
Figure~\ref{sw-set}, corresponding to a rest state (top row), eastward
equatorial jet (second row), westward equatorial jet (third row), and
three-jet pattern (fourth row).  As Figure~\ref{sw-low} demonstrates,
all aspects of the equilibrated, steady-state flow field---including
the spatial pattern of the equilibrated geopotential and the
zonal-mean zonal winds---are essentially identical regardless of the
initial condition used.  This final state consists
of standing, planetary-scale Rossby and Kelvin waves; two anticyclones
straddle the equator on the dayside and two cyclones straddle it on
the nightside \citep[see][]{showman-polvani-2011}.  
Because of the low forcing amplitude, the
equilibrated zonal-mean zonal wind is weak---corresponding to an
equatorial superrotating flow with a speed of only
$0.01\rm\,m\,s^{-1}$.  Note that all models equilibrate to this
identical final jet profile despite the fact that the speed of the initial jet,
$\pm1\rm\,km\,s^{-1}$, exceeds that of the equilibrated jet by a
factor of $10^5$.  

Figure~\ref{sw-high} illustrates an example of a high-amplitude model,
where $\Delta h_{\rm eq}/H=0.5$, $\tau_{\rm rad}=0.1\rm\,day$, and
$\tau_{\rm drag}\to\infty$ (meaning there is no explicit large-scale
drag in the upper layer; as described in \citet{showman-polvani-2011},
such a model still equilibrates because of interactions with the
quiescent lower layer).   Again, the four models in Figure~\ref{sw-high}
are integrated from the four initial conditions shown in
Figure~\ref{sw-set}, corresponding to a rest state (top row), eastward
equatorial jet (second row), westward equatorial jet (third row), and
three-jet pattern (fourth row).  All of the models equilibrate
to the same final state, with significant day-night differences in
geopotential and an overall pattern of eastward-equatorward phase
tilts, particularly on the dayside, which is the result of the
standing, planetary-scale Rossby and Kelvin waves.  Because of the
large forcing amplitude, short $\tau_{\rm rad}$, and absence of
large-scale drag, the zonal-mean zonal wind equilibrates to fast speeds
of $1\rm\,km\,s^{-1}$ in the core of the superrotating equatorial
jet that emerges (Figure~\ref{sw-high}).  Again, we emphasize that the
speed and amplitude of this equilibrated jet is totally independent
of whether the initial condition contained an eastward jet, a westward
jet, multiple jets, or no jets at all.

\begin{figure}[!h]
\centering
\includegraphics[scale=0.47, angle=0]{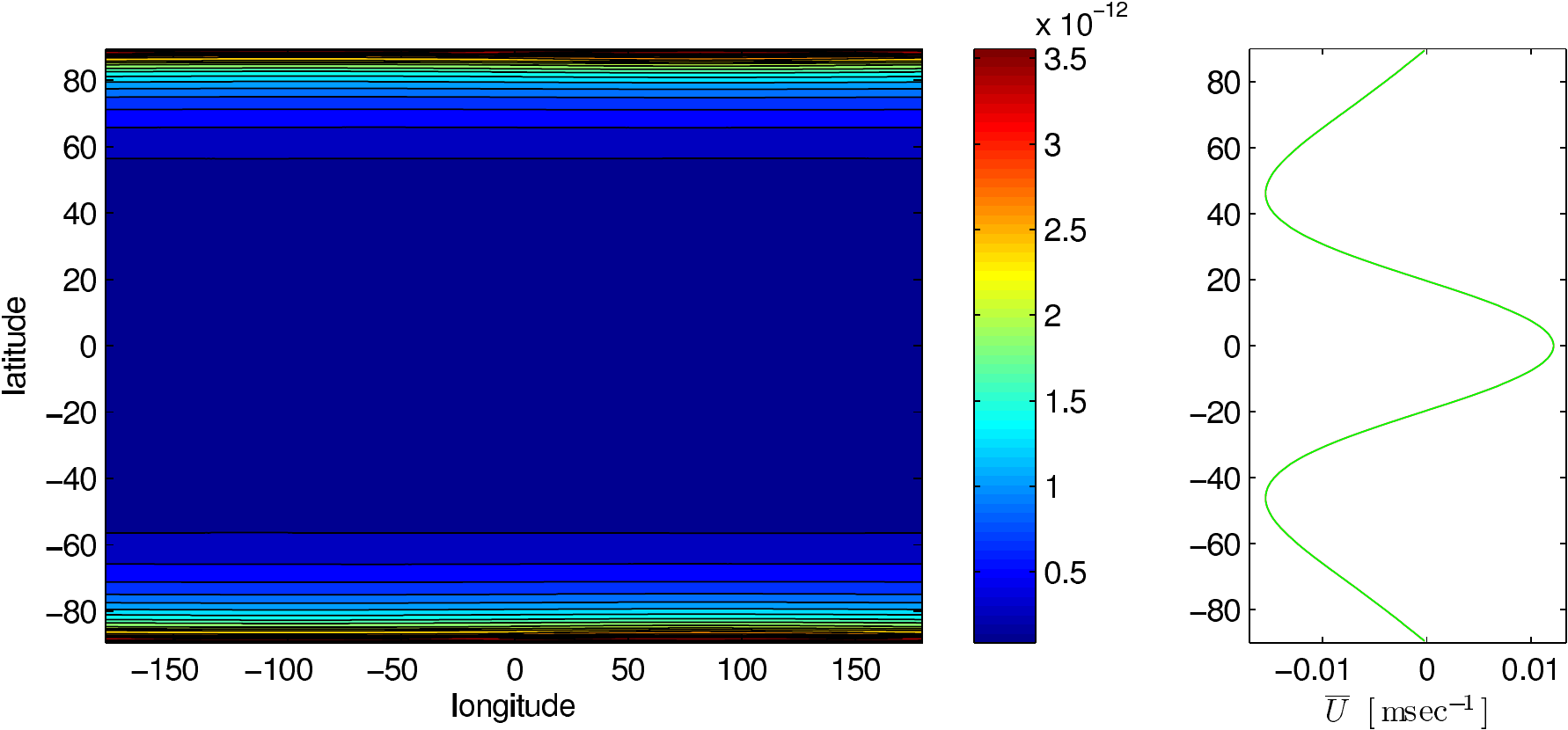} 
\includegraphics[scale=0.47, angle=0]{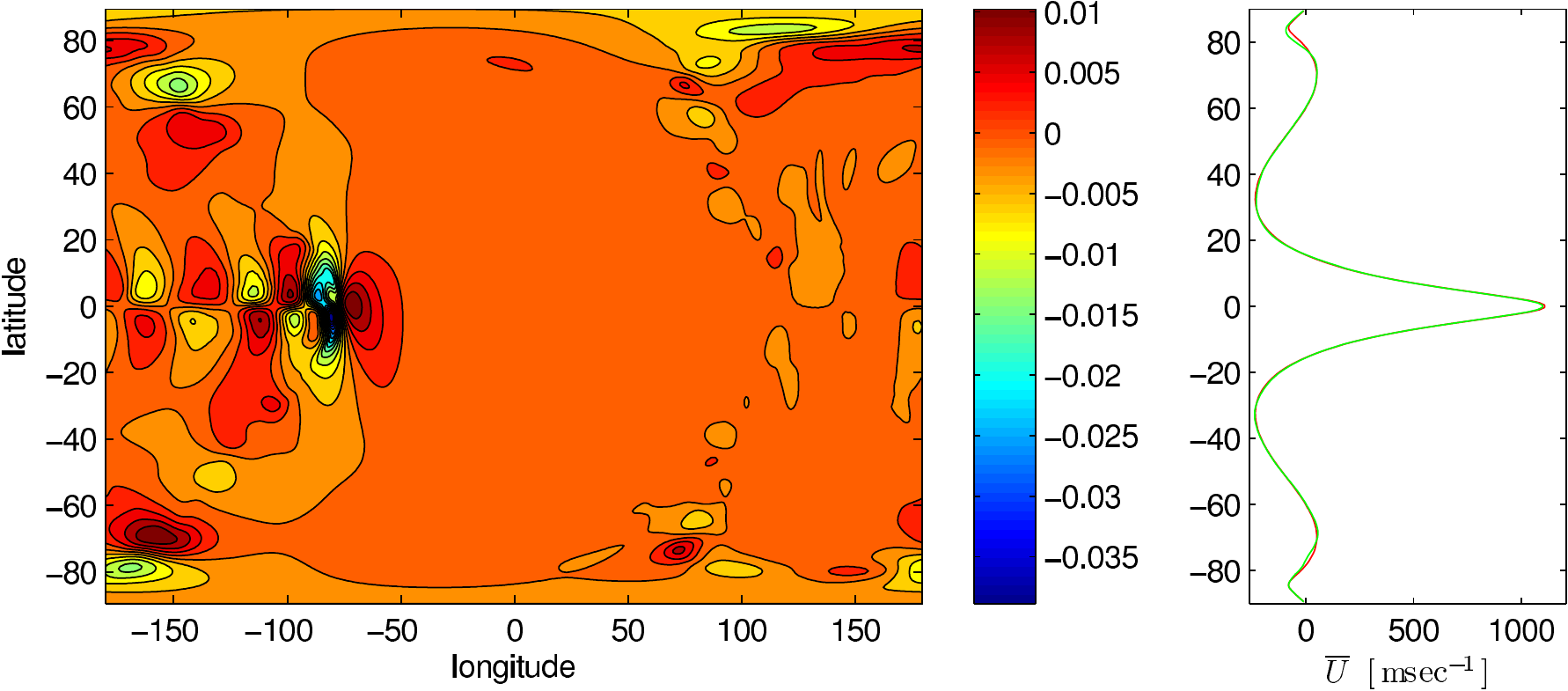}
\caption{Differences between the models in Figures~\ref{sw-low} and
  \ref{sw-high}.  {\it Left:} Fractional differences in geopotential
  ($gh$) between two otherwise identical models initialized with
  different initial conditions.  Top left plots $(gh_{\rm SW2} -
  gh_{\rm SW3})/gh_{\rm SW2}$, whereas bottom left plots $(gh_{\rm
    SW5} - gh_{\rm SW8})/gh_{\rm SW5}$, both at a specific instant in
  time (100 days in the top row, 1000 days in the bottom row).  Here,
  $gh_{\rm X}$ refers to the geopotential of model X. {\it Right}:
  Zonal-mean zonal wind from the two models shown in the corresponding
  left panels.  Top right shows SW2 (green curve) and SW3 (red curve)
  at 100 days while bottom right shows SW5 (green curve) and SW8 (red
  curve) at 1000 days.  The figure demonstrates that low-amplitude
cases (SW2 and SW3) converge to the identical steady state, with
fractional differences of $\sim$$10^{-12}$ or less; high-amplitude
cases, however, converge to states that, at any given instant, can
differ at specific points by up to $\sim$2\%, although the zonal-mean
zonal winds are still nearly identical.}
\label{sw-difference}
\end{figure}

Figure~\ref{sw-difference} further quantifies the similarity between
the final states of otherwise identical models initialized with
differing initial conditions.  The left panels show the differences in
the geopotential, at a given time, in the equilibrated states of two
models integrated with identical forcing parameters but differing
initial conditions, i.e., $[gh_{\rm model\,a}(\lambda,\phi, t_1) -
  gh_{\rm model\,b}(\lambda,\phi, t_1)]/gh_{\rm
  model\,a}(\lambda,\phi, t_1)$), where ``model a'' and ``model b''
are the two models being compared, and $t_1$ is some late time after
the runs are equilibrated.  The right panels overplot the zonal-mean
zonal wind for these two models in red and green.  The top row
represents the differences between two low-amplitude models (SW2 and
SW3) while the bottom row shows the differences between two
high-amplitude cases (SW5 and SW8).

When the forcing amplitude is low and the solutions are steady, the
final solutions are identical, in a point-to-point sense, to a
precision of literally $\sim$$10^{-12}$--$10^{-13}$ (top row of
Figure~\ref{sw-difference}).  Fractional differences are a
few$\,\times10^{-13}$ over most of the globe but rise to a
few$\,\times10^{-12}$ in a few localized regions (particularly near
the poles).  These miniscule differences are numerical, resulting from
a combination of roundoff and discretization error, and indicate that,
for all practical purposes, the solutions of these different models
are truly identical despite the vastly different initial conditions.
The equilibrated zonal-mean zonal wind profiles are likewise so
similar that the red curve is precisely covered by the overlying green
curve (top right panel of Figure~\ref{sw-difference}).

\begin{figure}[!h]
\centering
\includegraphics[scale=0.47, angle=0]{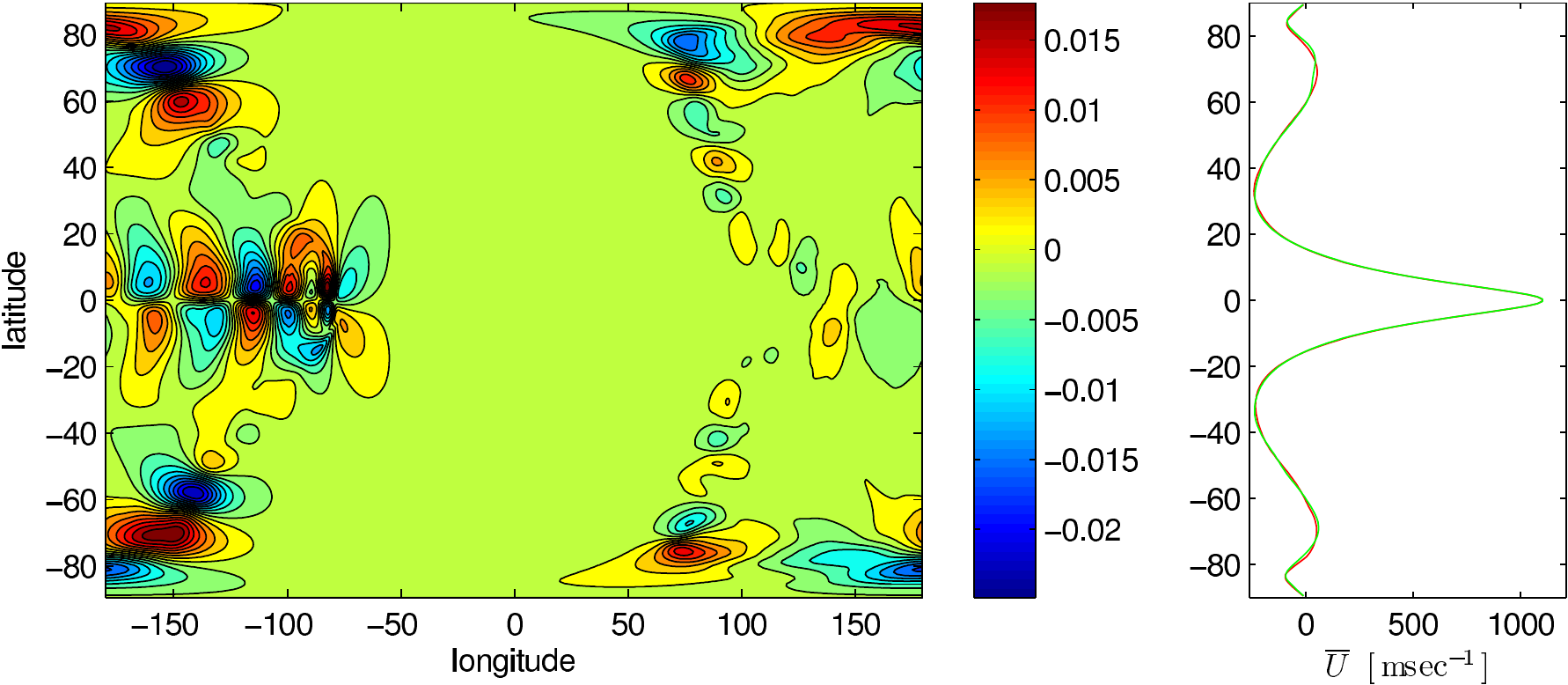}  
\caption{Differences between a given high-amplitude shallow-water model, SW6,
at two different points in time (900 and 1000 days), after the
models have reached statistical steady state.   Left panel shows the
fractional geopotential differences, i.e., $(gh_{\rm 900 days} -
gh_{\rm 1000 days})/gh_{\rm 900 days}$, whereas right panel shows the
zonal-mean zonal wind at 900 days (green curve) and 1000 days
(red curve).  These temporal differences within a given model
are similar to the model-model differences shown in the bottom
row of Figure~\ref{sw-difference}, demonstrating that those
differences result from time variability rather than sensitivity
of the statistical steady state to initial conditions.}
\label{sw-time}
\end{figure}

Although Figure~\ref{sw-high} demonstrates that high-amplitude models
likewise equilibrate to nearly identical states, tiny differences
become apparent when one compares in more detail.  The bottom row of
Figure~\ref{sw-difference} quantifies these differences.  In this
case, the fractional point-to-point differences {\it at a given time}
between otherwise identical models initialized with differing initial
conditions (in this case SW5 and SW8) reaches $\sim$1\% in specific
regions, particularly on the nightside.  One can ask whether this is a
true difference in the statistical steady states between these models
or whether it rather results from time variability that might induce a
slight randomness {\it around a single} statistical steady
state.  To address this question, Figure~\ref{sw-time} shows the
differences between two different snapshots at different times within
a {\it given} model integration.  In the bottom row, these differences
are likewise seen to reach $\sim$1\%, with a spatial pattern extremely
similar to that seen in the bottom row of Figure~\ref{sw-difference}.
This comparison indicates that the model-model differences shown in
Figure~\ref{sw-difference} are the simple result of time variability
and {\it not} the result of any fundamental sensitivity of the
statistical steady state to initial condition.  Indeed, the
differences between the {\it time-averages} of the same two models are
much smaller than the differences between their instantaneous
snapshots displayed in Figure~\ref{sw-difference} (not shown),
confirming that the statistical steady states are essentially
identical for these models.  We also note that, even at a given time,
the two models have nearly identical zonal-mean zonal wind profiles
(Figure~\ref{sw-difference}, lower right panel), indicating that even
the instantaneous point-to-point variability has little effect on the
zonal-mean state.

We emphasize that the results shown are not specific to the 
particular parameters illustrated in the figures but rather
are general.  The wide variety of models we have performed all
confirm the essential point made here, namely, the insensitivity
of this forced shallow-water model to initial conditions.

\section{three-dimensional model}
\label{3D model}

\subsection{ Model description}

We now consider 3D models of the atmospheric circulation.  As in
most previous investigations, we adopt the primitive equations.
We solve the equations using the MITgcm, which is a state-of-the-art
circulation model \citep{adcroft-etal-2004} that
\citet{showman-etal-2009} adapted for application to hot Jupiters.
The horizontal momentum, vertical momentum, continuity, and thermodynamic energy equations are, using pressure as a vertical coordinate,
\begin{equation}
{d{\bf v}\over dt}= -\nabla \Phi - f {\bf k}\times {\bf v} + {\cal D}_{\bf v}
\label{momentum}
\end{equation}
\begin{equation}
{\partial \Phi\over \partial p}=-{1\over\rho}
\label{hydrostatic}
\end{equation}
\begin{equation}
\nabla \cdot {\bf v} + {\partial \omega\over\partial p}=0
\label{continuity}
\end{equation}
\begin{equation}
{d T\over dt} = {q\over c_p} + {\omega \over \rho c_p}
\label{energy}
\end{equation}
where ${\bf v}$ is the horizontal velocity on constant-pressure
surfaces, $\omega\equiv dp/dt$ is the vertical velocity in pressure
coordinates, $\Phi$ is the gravitational potential on
constant-pressure surfaces, $f\equiv 2\Omega\sin\phi$ is the Coriolis
parameter, $\Omega$ is the planetary rotation rate, ${\bf k}$ is the
local vertical unit vector, $q$ is the thermodynamic heating rate
($\W\kg^{-1}$), and $T$, $\rho$, $c_p$ are the temperature, density,
and specific heat at constant pressure.  $\nabla$ is the horizontal
gradient evaluated on constant-pressure surfaces, and
$d/dt=\partial/\partial t + {\bf v}\cdot \nabla + \omega
\partial/\partial p$ is the material derivative (including
curvature terms in spherical geometry).  The term $\cal D_{\bf v}$
is a velocity damping term, including a Shapiro filter to
maintain numerical stablity (which has only a small effect on the
large-scale flow), and optionally, an explicit large-scale
frictional drag term (see below).  Eq~(\ref{energy}) is
actually solved in an alternate form,
\begin{equation}
{d\theta\over dt}={\theta\over T}{q \over c_p} 
\label{energy2}
\end{equation}
where $\theta=T(p/p_0)^{\kappa}$ is the potential temperature (a
measure of entropy), $\kappa$ is the ratio of gas constant to specific
heat at constant pressure, and $p_0$ is a reference pressure (note
that the dynamics are independent of the choice of $p_0$).  The
dependent variables ${\bf v}$, $\omega$, $\Phi$, $\rho$, $\theta$, and
$T$ are functions of longitude $\lambda$, latitude $\phi$, pressure
$p$ and time $t$.

\begin{figure}[ht!]
\centering
\includegraphics[scale=0.46, angle=0]{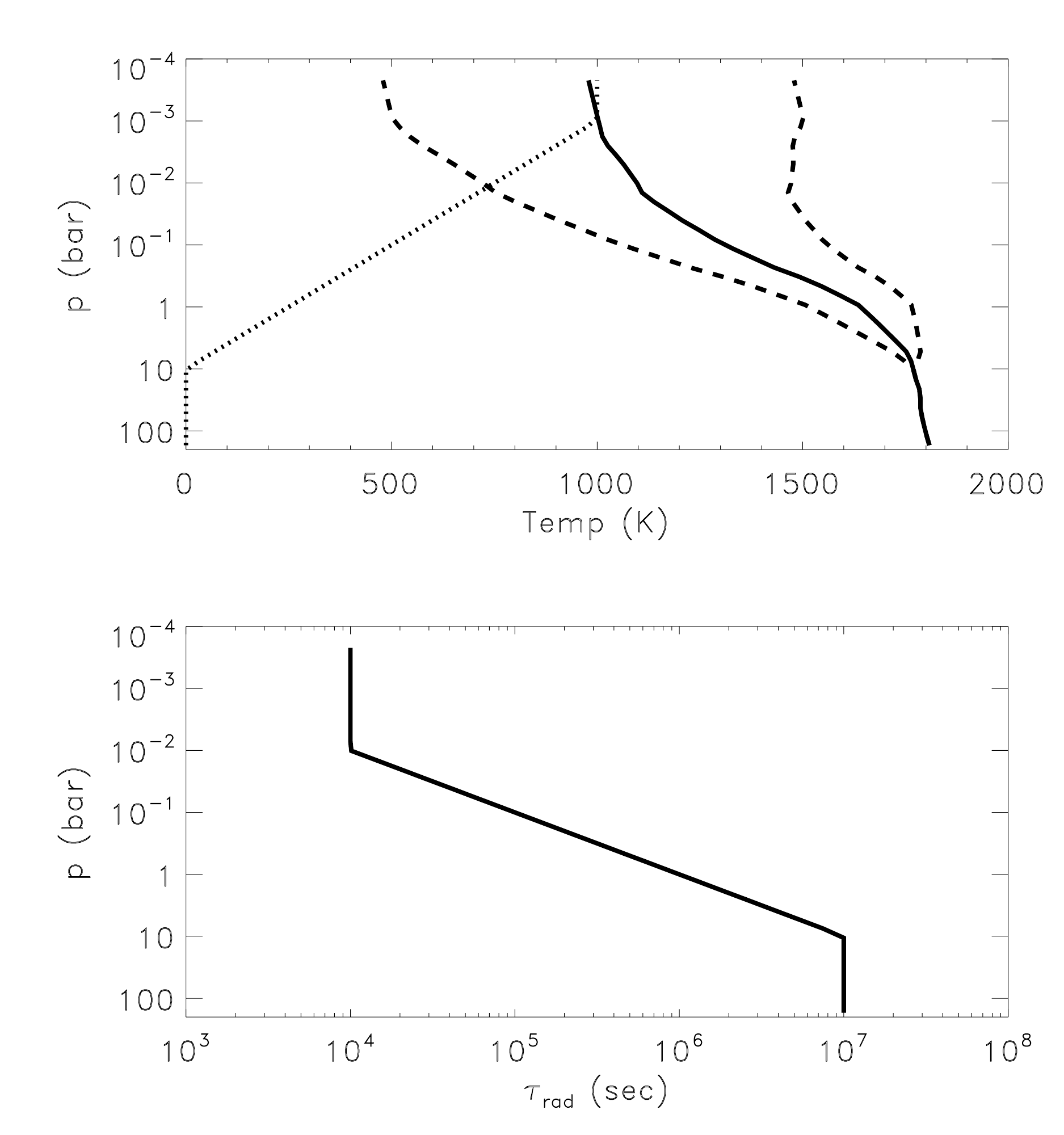} 
\caption{Forcing profiles used in the Newtonian heating/cooling
  scheme.  {\it Top:} Radiative-equilibrium temperature profiles.
  Dashed curves show the radiative equilibrium temperature profile for
  the nightside ($T_{\rm night,eq}$, left curve) and substellar
  point (right curve).  The solid curve shows $T_{\rm Iro}(p)$.  The
  dotted curve shows the substellar-nightside radiative-equilibrium
  temperature difference, $\Delta T_{\rm eq}(p)$, assuming the
  temperature difference at the top, $\Delta T_{\rm eq,top}$, is
  $1000\rm\,K$.  {\it Bottom:} Radiative time constant, $\tau_{\rm
    rad}$, versus pressure.}
\label{rad}
\end{figure}

\citet{showman-etal-2009} coupled the MITgcm to the multi-stream
radiative transfer model of \citet{marley-mckay-1999}, which allows
for accurate calculation of heating rates when the atmospheric 
composition and opacities are specified.  In the present context,
however, our goal is to characterize the sensitivity to initial
conditions in the clearest possible context, and so rather
than using this coupled model, we specify the radiative heating/cooling
using a Newtonian cooling scheme, which relaxes the temperature
toward a specified radiative-equilibrium temperature over a 
specified time constant: 
\begin{equation}
\label{equation:Cooling}
\frac{q}{c_p} = - \frac{T(\lambda,  \phi, p ,t) - T_{\rm eq}(\lambda, \phi,p)}{\tau_{\rm rad} (p)}.
\end{equation}
The Newtonian cooling scheme has been
widely used in exoplanet studies \citep{showman-guillot-2002,
cooper-showman-2005, showman-etal-2008a, menou-rauscher-2009, 
rauscher-menou-2010, perna-etal-2010, thrastarson-cho-2010,
heng-etal-2011}.


\begin{figure*}
\centering
\includegraphics[scale=0.7, angle=0]{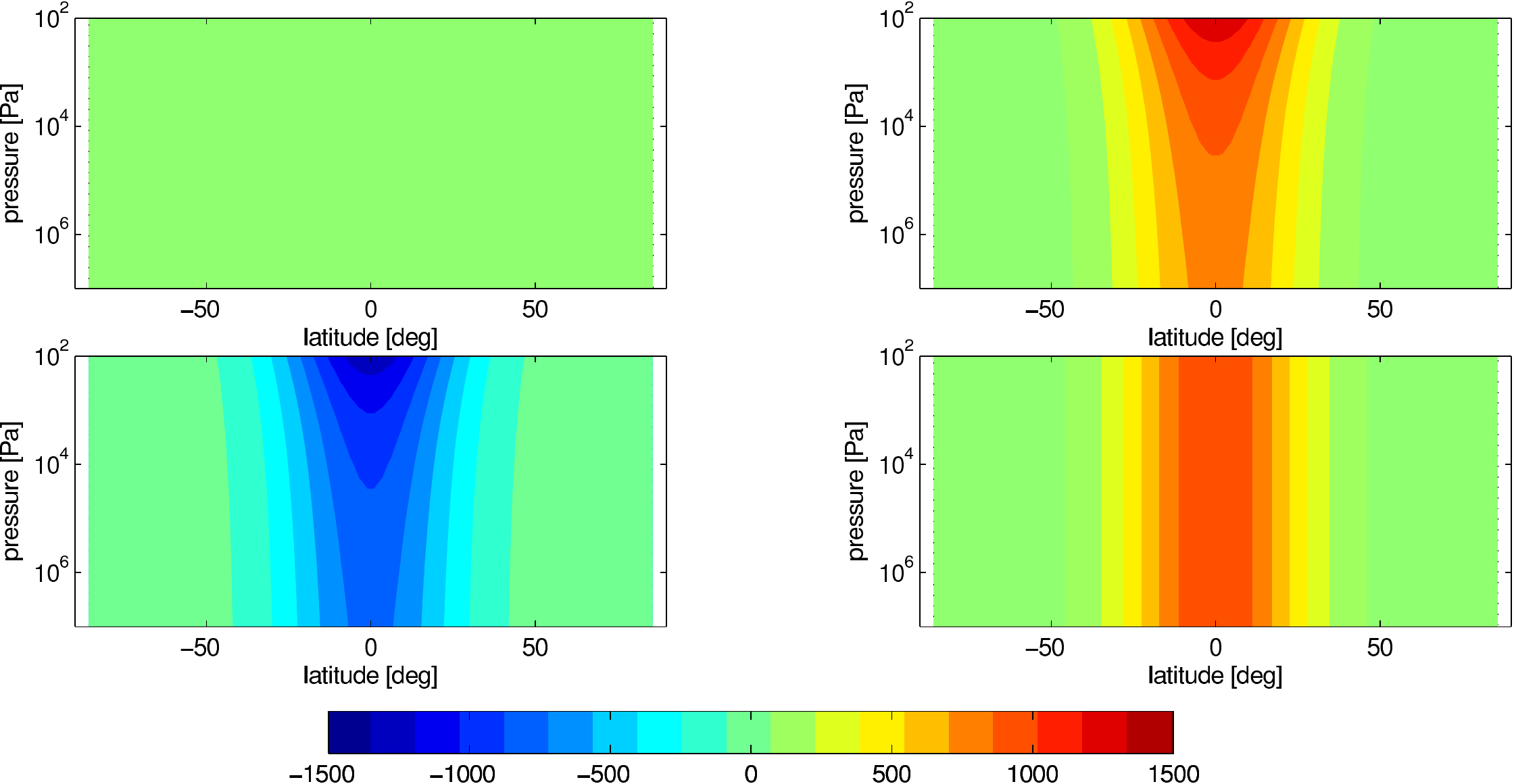} 
\caption{Initial jet profiles adopted in our 3D models.  Plotted is zonal-mean zonal wind
of the initial condition, in $\rm m\,s^{-1}$, versus latitude and pressure.  Initial
conditions correspond to a rest state (upper left), eastward decaying jet (upper right),
westward decaying jet (lower left), and eastward barotropic jet (lower right).
We also tried westward barotropic jets in some models (not shown).}
\label{gcmjet}
\vskip 10pt
\end{figure*}

The radiative-equilibrium temperature, $T_{\rm eq}(\lambda,\phi,p)$,
is defined as 
\begin{equation}
T_{\rm eq}(\lambda,\phi,p) = 
\begin{cases}
   T_{\rm night,eq}(p) + \Delta T_{\rm eq}(p)\cos\lambda\cos\phi,   &{\rm dayside};\\
   T_{\rm night,eq}(p), &{\rm nightside}
\label{teq}
\end{cases}
\end{equation}
where $T_{\rm night,eq}(p) + \Delta T_{\rm eq}$ is the
radiative-equilibrium temperature at the substellar point and $\Delta
T_{\rm eq}(p)$ is the difference in radiative-equilibrium temperature
between the substellar point and the nightside.  As written this
expression takes the substellar point to be at longitude and latitude
of $0^{\circ},0^{\circ}$.  To specify the nightside profile, we
further define $T_{\rm night,eq}(p) = T_{\rm Iro} - {\Delta T_{\rm
    eq}\over 2}$, where $T_{\rm Iro}(p)$ is the one-dimensional
radiative-equilibrium temperature profile from \citet{iro-etal-2005}.
These definitions then imply that the substellar radiative-equilibrium
temperature profile is $T_{\rm Iro} + {\Delta T_{\rm eq}\over 2}$.
Since radiative heating is strong at the top and weak at the bottom,
it is important that $\Delta T_{\rm eq}(p)$ decreases with increasing
pressure.  For computational simplicity we specify $\Delta T_{\rm
  eq}(p)$ as a piecewise-continuous analytical function that is a
constant, $\Delta T_{\rm eq,top}$, at pressures less than $p_{\rm
  eq,top}$; is zero at pressures exceeding $p_{\rm eq,bot}$; and
varies linearly with log-pressure in between.  For the models
described in this paper, we take $\Delta T_{\rm eq,top}=1000\rm\,K$,
$p_{\rm eq,top}=10^{-3}\rm\,bar$, and $p_{\rm eq,bot}=10\rm\,bar$;
note that our key result---namely, insensitivity to initial
conditions---does not depend on these precise values.  The nightside
and substellar radiative equilibrium profiles, as well as $T_{\rm
  Iro}$ and $\Delta T_{\rm eq}(p)$, are shown in Figure~\ref{rad}.

Likewise, the radiative time constant is expected to be a strong
function of pressure, being short at the top and long at the bottom
\citep{iro-etal-2005, showman-etal-2008a}.  For computational
simplicity, we here assume that $\tau_{\rm rad}$ is a function of
pressure alone, and we again define a piecewise-continuous analytic
function that allows such a downward-increasing behavior: we take
$\tau_{\rm rad}(p)$ to be a constant, $\tau_{\rm rad,top}$, at $p\le
p_{\rm rad,top}$; another constant, $\tau_{\rm rad,bot}$, at $p\ge
p_{\rm rad,bot}$; and we assume that $\log\tau_{\rm rad}$ varies
linearly with $\log p$ in between.  In this paper, we take $p_{\rm
  rad,top}=10^{-2}\rm\,bar$ and $p_{\rm rad,bot} = 10\rm\,bar$.  We
explore several values for $\tau_{\rm rad,top}$ and $\tau_{\rm
  rad,bot}$ in different models, with $\tau_{\rm rad,top}$ generally
chosen to be short and $\tau_{\rm rad,bot}$ chosen to be long.  Again,
our key results are not dependent on the precise values. The profile
of $\tau_{\rm rad}$ for one such model is shown in Figure~\ref{rad}.

In our models, we also include a simple frictional drag scheme near
the bottom of the domain.  This might crudely represent the effects of
``magnetic drag'' associated with the partial ionization expected at
temperatures exceeding $\sim$$1500\rm\,K$ \citep{perna-etal-2010,
  menou-2012}, which occur in our model at pressures exceeding
$\sim$$1\rm\,bar$ (see Figure~\ref{rad}).  From a more practical
perspective, such frictional drag also forces the flow to equilibrate
in a reasonable integration time.  When studying sensitivity to
initial conditions, it is particularly important to ensure that the
models have reached equilibrium, and this is aided by including such a
drag scheme.  The drag is introduced on the righthand side of
Equation~(\ref{momentum}) and takes the form $-k_v{\bf v}$, where
$k_v(p)$ is a pressure-dependent drag coefficient.  The drag
coefficient is zero at pressures less than $p_{\rm drag,top}$ and
equal to $k_F(p-p_{\rm drag,top})/ (p_{\rm bot}-p_{\rm drag,top})$ at
$p\ge p_{\rm drag, top}$, where $p_{\rm drag,top}$ is the lowest
pressure of the region experiencing drag, $p_{\rm bot}$ is the mean
pressure at the bottom of the domain, and $k_F$ is a constant
\citep[this formulation of drag is extremely similar to that of][]
{held-suarez-1994}.  This formulation implies that
the drag coefficient increases linearly with pressure from zero at
$p_{\rm drag,top}$ to $k_F$ at the bottom of the domain.  Motivated
by expectations that magnetic drag is most important only at temperatures
exceeding $\sim$$1500\rm\,$K, we take $p_{\rm drag,top} = 1\rm\,bar$
in most models, although we also explore values of $10\rm\,bar$
to determine the sensitivity to drag scheme.  The qualitative structure
of the equilibrated dynamical
state is not strongly sensitive to $k_F$; here, we
explore values of $0.1$ and $0.01\rm\,day^{-1}$, implying characteristic
drag timescales of 10 and 100 days near the bottom.

Overall, our choices of $\Delta T_{\rm eq}(p)$, $\tau_{\rm rad}(p)$
and drag scheme described above are motivated by three overarching
goals: (1) to ensure that the radiative heating/cooling rates
(expressed in $\rm K\,s^{-1}$) are large at the top but decrease
rapidly with increasing pressure to very small values at the bottom,
as expected on real hot Jupiters; (2) to produce equilibrated
circulation patterns qualitatively resembling those from models that
couple the dynamics to radiative transfer
\citep[i.e.,][]{showman-etal-2009, heng-etal-2011b, rauscher-menou-2012,
perna-etal-2012}, and (3) to ensure that the models
equilibrate in finite time, as necessary to test sensitivity to
initial conditions and to survey the parameter space.  The first and
second criteria generally lead to choices of $\tau_{\rm rad,top}=10^4$
or $10^5\rm\,s$ and $\tau_{\rm rad,bot}\gtrsim 10^6\rm\,s$, while the
third criterion suggests $\tau_{\rm rad,bot}\lesssim 10^8\rm\,s$
and $p_{\rm drag,top}\lesssim 10\rm\,$bars.   We note that our
fomulation differs significantly from that of \citet{thrastarson-cho-2010},
who choose $\Delta T_{\rm eq}$ and $\tau_{\rm rad}$ to be independent
of pressure; although this assumption has the advantage of simplicity, it
fails to satisfy criteria (1) and (2).

\begin{table*}
\centering
\caption{The properties of some 3D runs}
\begin{tabular}{|c|c|c|c|c|c|}
\hline
\hline
Name & Initial condition &$\tau_{\rm rad,top}\rm\,(s)$ & $\tau_{\rm rad,bot}\rm\,(s)$  & $kF\rm\,(day^{-1})$  &$p_{\rm drag,top}\rm\,(bar)$)\\
\hline
GCM1 & Rest state & $10^4$   &     $10^6$    &   $0.1$  & 1 \\    
GCM2 & Eastward decaying jet & $10^4$   &     $10^6$    &   $0.1$ & 1  \\  
GCM3 & Westward decaying jet & $10^4$   &     $10^6$    &   $0.1$  & 1\\   
GCM4 & Eastward barotropic jet & $10^4$   &     $10^6$    &   $0.1$  & 1\\   
GCM5 & Rest state  & $10^5$     &    $10^6$    &   $0.1$  & 1\\   
GCM6 & Eastward decaying jet & $10^5$     &    $10^6$    &   $0.1$ & 1 \\  
GCM7 & Westward decaying jet & $10^5$     &    $10^6$    &   $0.1$ & 1 \\  
GCM8 & Eastward barotropic jet & $10^5$     &    $10^6$    &   $0.1$  & 1\\  
\hline
\end{tabular} 
\label{tab2}
\end{table*}

Following \citet{thrastarson-cho-2010}, as well as our shallow-water models
from Section~\ref{2D model}, we initialize most of our 3D models with
a zonally symmetric zonal flow whose latitude dependence is given
by Equation~(\ref{initial-jet}) with $\sigma=\pi/9$ and $U=0$ 
(corresponding to a rest state), $1\rm\,km\,s^{-1}$
(corresponding to an eastward equatorial jet), or $-1\rm\,km\,s^{-1}$
(corresponding to a westward equatorial jet).  In some cases, we
assume this initial jet to be independent of pressure, while in
others, we allow the jet to decay with pressure by multiplying
the right side of Equation~(\ref{initial-jet}) by the function
$1.6 + \arctan[-(\log p - \log p_{\rm top})/\log p_{\rm top}]/1.6$, which
causes the jets to decay from a peak speed of $\sim$$1.6\rm\,km\,s^{-1}$
at the top to $\sim$$400\rm\,m\,s^{-1}$ at the bottom of the domain.
Figure~\ref{gcmjet} shows several of these initial conditions, laid
out in a format that we will repeat, for easy comparison, when
presenting results.

We adopt planetary parameters appropriate to hot Jupiters,
including specific heat $\rm
c_p=1.3\times10^4\J\kg^{-1}\,{\rm K}^{-1}$, specific gas constant
$3700\rm\,J\,kg^{-1}\,K^{-1}$, and
a rotation  period, gravity, and planetary radius of
$3.024\times10^{5}\,{\rm s}$,
$9.36\m\,{\rm s}^{-2}$, and $9.437\times10^7\,$m, respectively. 
These values are appropriate to HD 209458b, although we emphasize
that our results are not sensitive to the precise values,
and similar behavior would occur had choices appropriate to other
typical hot Jupiters been made instead.

\begin{figure*}
\includegraphics[scale=0.8, angle=0]{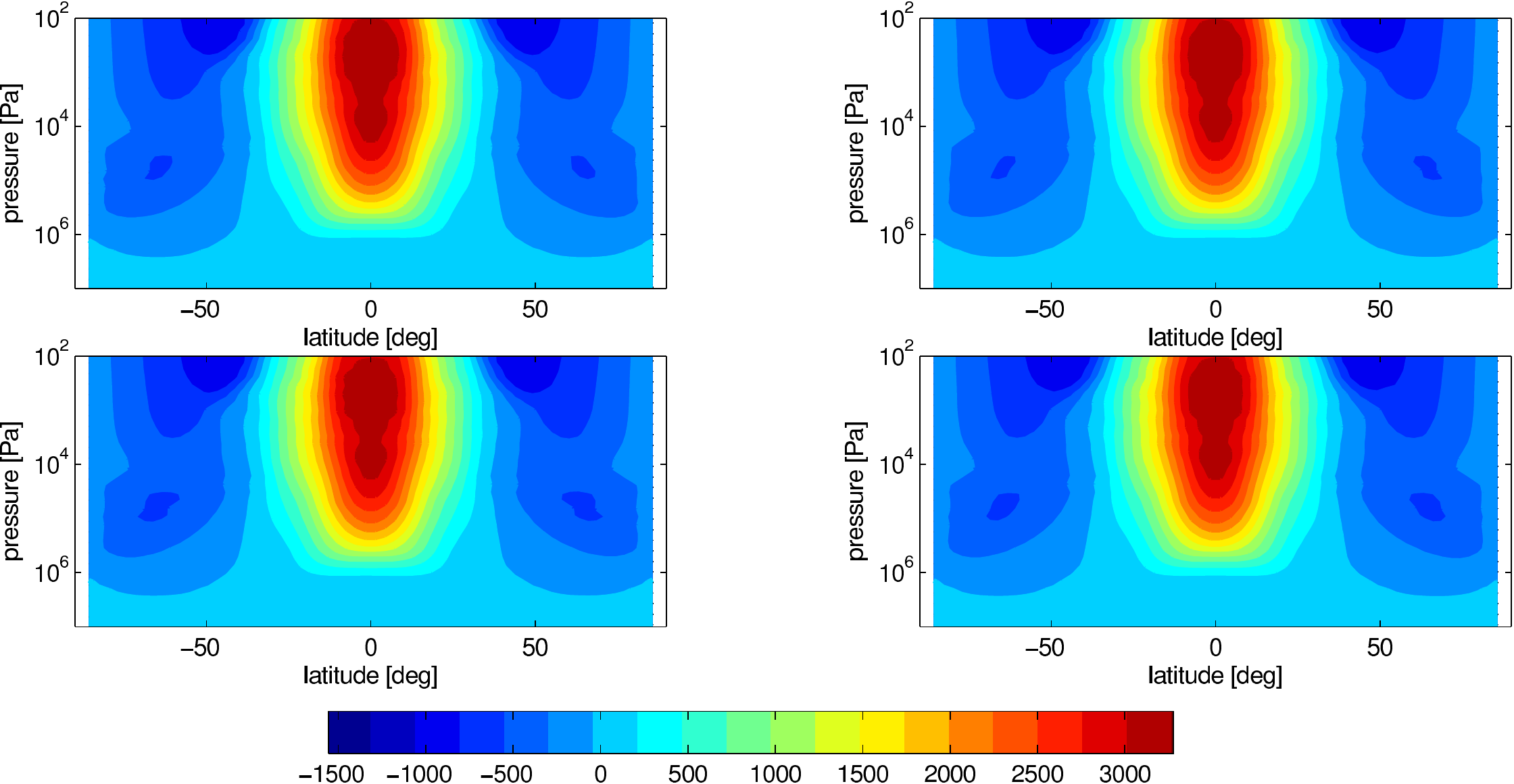} 
\caption{ Equilibrated zonal-mean zonal wind ($\rm m\,s^{-1}$) in four otherwise
  identical models with differing initial conditions.  These four
  models adopt the initial conditions corresponding to the equivalent
  panels of Figure~\ref{gcmjet}.  Upper left, upper right, lower left,
  and lower right panels show models GCM1, GCM2, GCM3, and GCM4,
  initialized from a rest state, an eastward decaying jet, a westward
  decaying jet, and an eastward barotropic jet, respectively.  The
  zonal averages shown here are time averaged from 694 to 1042 days.
  All four models adopt $\tau_{\rm rad,top}=10^4\rm\,s$, $\tau_{\rm
    rad,bot}=10^6\rm\,s$, $k_F=0.1\rm\,day^{-1}$, $p_{\rm
    drag,top}=1\rm\,bar$, and $\Delta T_{\rm eq,top}=1000\rm\,K$.  }
\label{u1}
\end{figure*}

The MITgcm solves the equations using a finite-volume discretization
on staggered Arakawa C grid \citep{arakawa-lamb-1977}.  Rather than
the standard longitude/latitude coordinate system, we solve the
equations using the cubed-sphere grid following
\citet{showman-etal-2009}. The horizontal resolutions is C32 in most
models, implying that each of the six ``cube faces" has a resolution
of $32\times32$ finite-volume elements, which is roughly equal to a
global resolution of $128\times64$ in longitude and latitude.
However, to ensure that our results are numerically converged and do
not depend on these numerical details, we also performed some models
at a resolution of C64 (i.e., $64\times64$ cells on each cube face,
corresponding to a global resolution of approximately $256 \times
128$) and a resolution of C128 (i.e., $128\times128$ on each cube
face, corresponding to a global resolution of approximately $512\times
256$).  The upper boundary is zero pressure and the bottom
boundary is an impermeable surface.  We adopt $N_L=40$ levels
in the vertical; the bottom $N_L-1$ levels are evenly spaced
in log-pressure between 200 bars at the bottom and 0.2 mbar
at the top; the top layer extends from a pressure of 0.2 mbar to zero.

\subsection{Results}

As before, we explored a variety of models, with differing values of
$\tau_{\rm rad,top}$, $\tau_{\rm rad,bot}$, $kF$, $p_{\rm drag,top}$ and
initial condition.  A small subset of these models, which are 
illustrated in subsequent figures, are shown in Table~\ref{tab2}.

\begin{figure*}
\includegraphics[scale=0.8, angle=0]{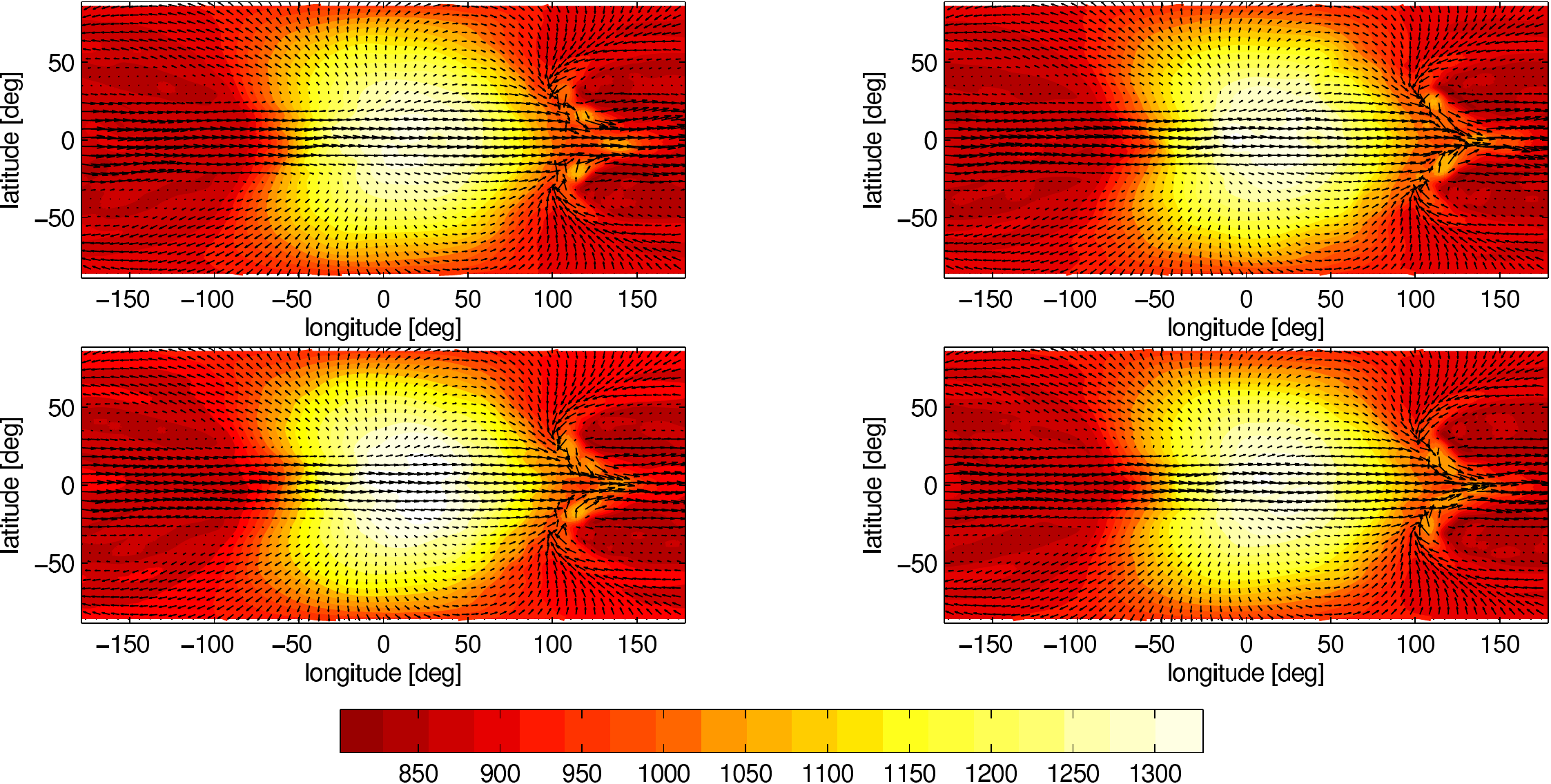}  
\caption{ Temperature (color scale, in K) and winds (arrows) for the
  same four models shown in Figure~\ref{u1}. This is at 30 mbar, which
  is near the infrared photosphere on a typical hot Jupiter.  Panels
  show models GCM1 (upper left), GCM2 (upper right), GCM3 (lower
  left), and GCM4 (lower right), initialized with a rest state,
  eastward decaying jet, westward decaying jet, and eastward
  barotropic jet, respectively.  These are snapshots at 1042 days.
  The final states are extremely similar.}
\label{temp1}
\end{figure*}

\begin{figure}
\includegraphics[scale=0.5, angle=0]{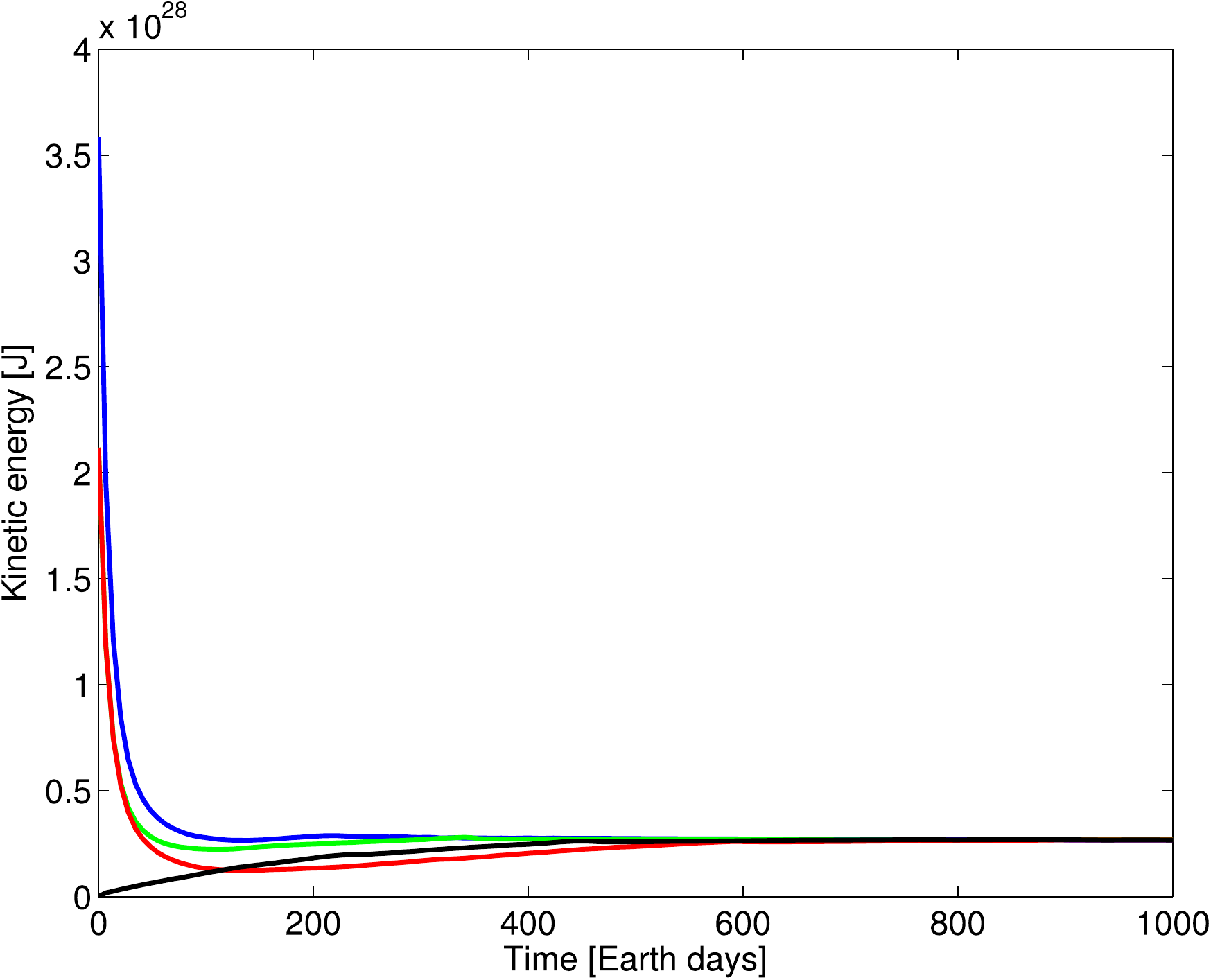}  
\caption{Total kinetic energy, integrated over the model domain,
for the four models GCM1 (black), GCM2 (green), GCM3 (red) and GCM4 (blue)
which are initialized with different initial conditions but are 
otherwise the same.
The initial kinetic energies differ greatly due to the differing
initial conditions, but the model all converge to the same final state,
independent of initial condition.}
\label{Ek}
\end{figure}

In agreement with our shallow-water results, we find that the final,
equilibrated state of our 3D models are independent of the initial
condition.  This is illustrated for a particular choice of forcing
parameters (corresponding to models GCM1 to GCM4) in Figures~\ref{u1}
and \ref{temp1}.  For easy comparison, the four panels in each of
these figures adopt the initial conditions of the corresponding panels
of Figure~\ref{gcmjet}---a rest state (top left panel), eastward
decaying jet (top right), westward decaying jet (bottom left), and
eastward barotropic\footnote{In this context, ``barotropic'' means
  that the horizontal wind speed is independent of pressure.} jet
(bottom right).  Figure~\ref{u1} shows the zonal-mean zonal wind,
while Figure~\ref{temp1} shows the temperature and two-dimensional
velocity pattern at a pressure of $30\rm\,mbar$, for these four models
after equilibrium has been reached.  As can be seen, all four models
exhibit extremely similar patterns of zonal wind, temperature pattern,
and two-dimensional velocity structure despite the differing initial
conditions.  The momentum fluxes caused by the day-night thermal
forcing drive a broad equatorial jet whose peak speeds exceed
$3\rm\,km\,s^{-1}$ (Figure~\ref{u1}).  The day-night temperature
differences exceed $\sim$$800\rm\,K$ at the top of the model and are
$500\rm\,K$ at 30 mbar (Figure~\ref{temp1}).  At low pressure, the
short radiative time constant---$10^4\rm\,s$ in these models---leads
to little longitudinal offset of the dayside hot region, although by
30 mbar the hot spot is displaced to the east of the substellar point
by $\sim$$20^{\circ}$ longitude. Figure~\ref{Ek} shows the total
kinetic energy over time, integrated over the entire domain, for these
four models; the initial kinetic energies differ because of the
differing initial jet profiles, but the models all converge to the same
kinetic energy over time.  Clearly, the models have lost memory of the
initial condition and have all converged to the identical statistical
steady state.

\begin{figure*}[htb!]
\includegraphics[scale=0.8, angle=0]{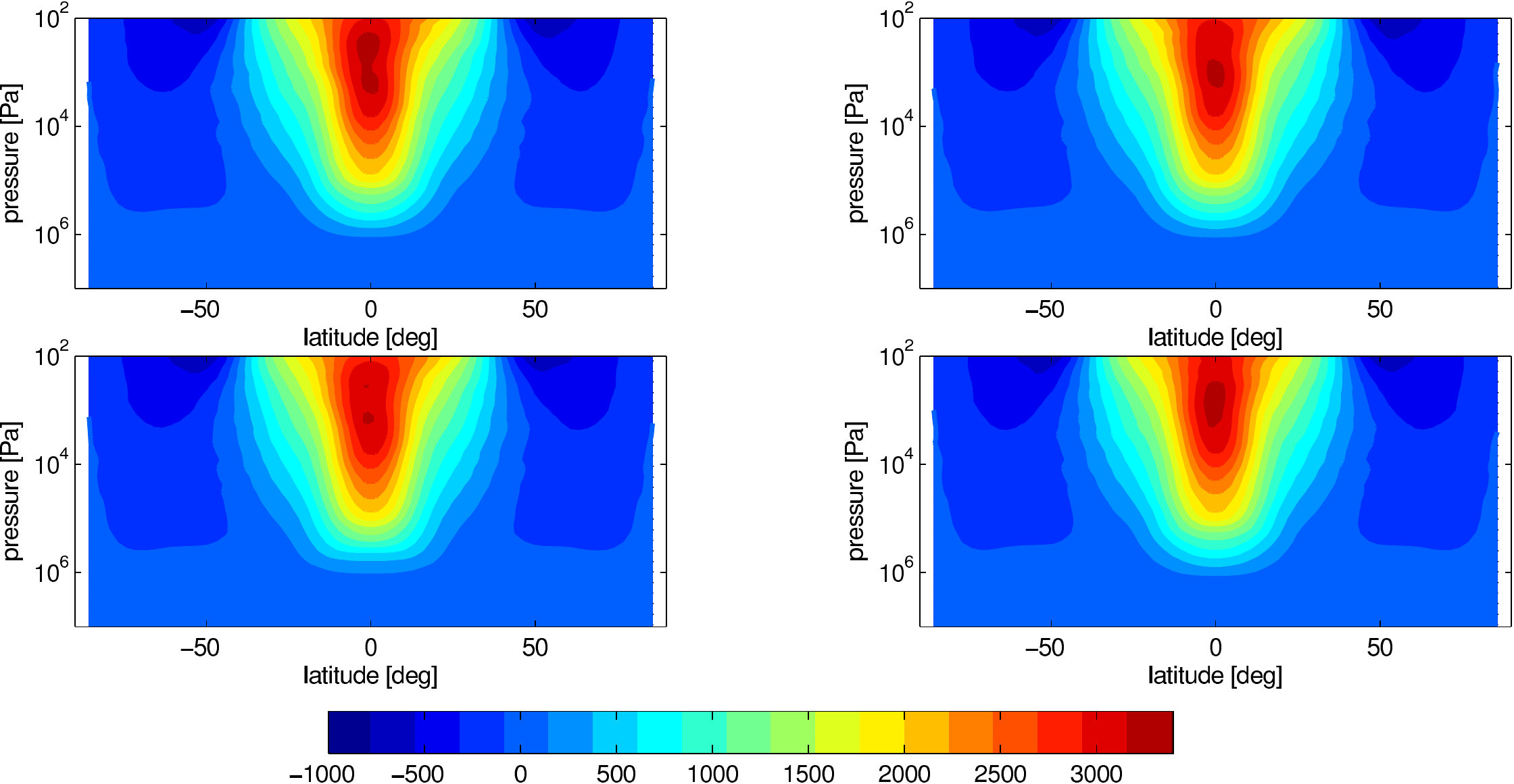}  
\caption{ 
Zonal-mean zonal wind ($\rm m\,s^{-1}$) in four otherwise identical models with differing
initial conditions.   These four models adopt the initial conditions
shown in Figure~\ref{gcmjet}.  Upper left, upper right, lower left,
and lower right panels show models GCM5, GCM6, GCM7, and GCM8, initialized
from a rest state, an eastward decaying jet, a westward decaying jet,
and an eastward barotropic jet, respectively.  Unlike Figure~\ref{u1},
this figure shows snapshots at 1042 days, giving a sense of the
instantaneous variability in the zonal-mean zonal wind.  In a time
average, this variability averages out and the different models
would look even more similar.
}
\label{u2}
\end{figure*}

\begin{figure*}[htb!]
\includegraphics[scale=0.8, angle=0]{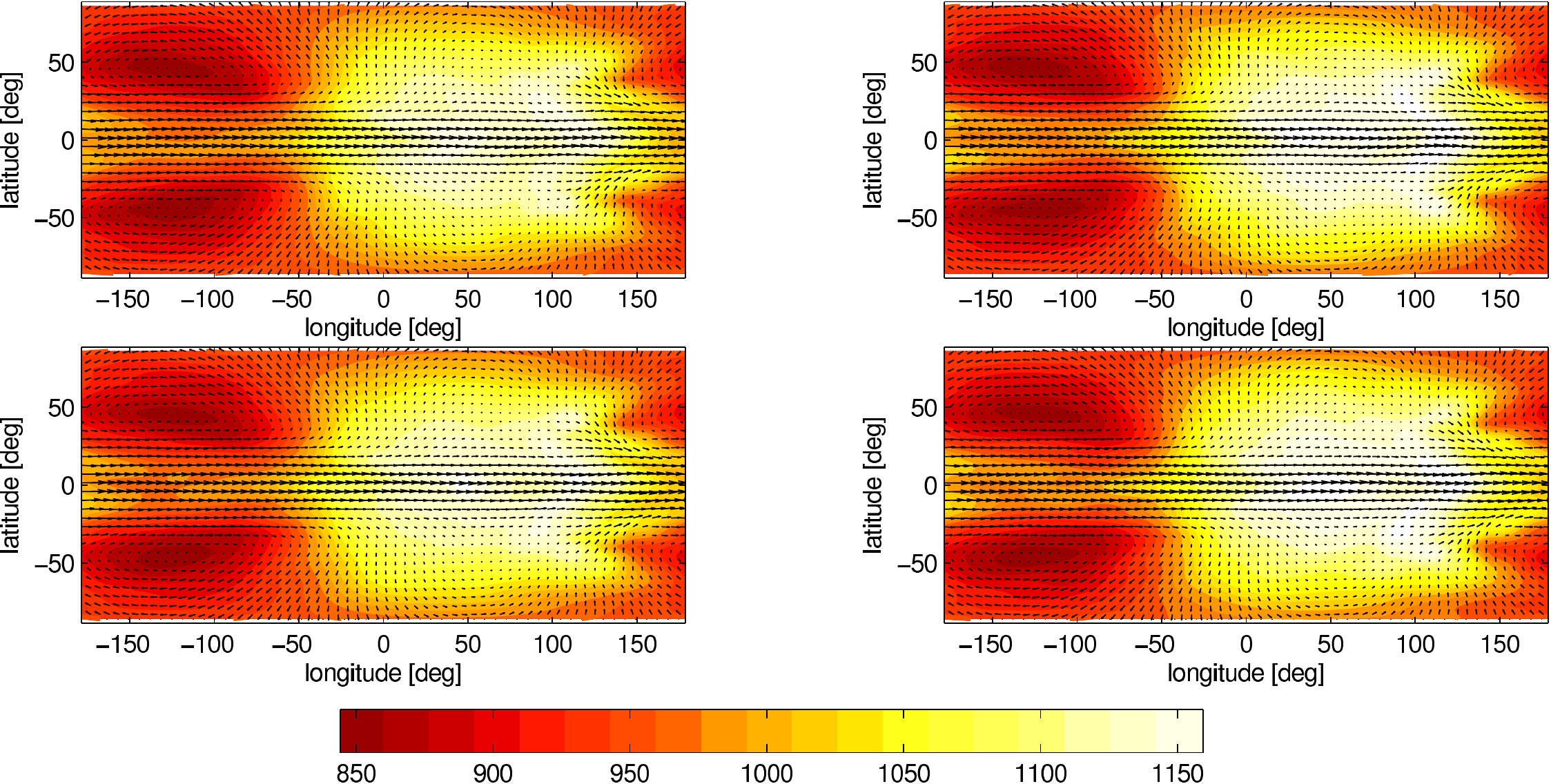}  
\caption{ 
Temperature at 30 mbar (color scale, in K) and winds (arrows) for the same four
models shown in Figure~\ref{u2}. Panels 
show models GCM5 ({\it top left}), GCM6 ({\it top right}), GCM7 
({\it bottom left}), and GCM8 ({\it bottom right}), initialized
with a rest state, eastward decaying jet, westward decaying jet,
and eastward barotropic jet, respectively.  These are snapshots 
at 1042 days, and the slight differences between panels gives
a sense of the time variability.  In a time average, the different
models look even more similar.
}
\label{temp2}
\end{figure*}

The lack of sensitivity to initial conditions is demonstrated with
another example in Figures~\ref{u2} and \ref{temp2}, which show models
with a longer radiative time constant of $10^5\rm\,s$ in the upper
part of the domain.  As before, the figures depict the zonal-mean
zonal wind (Figure~\ref{u2}) and the temperature and velocity patterns
at 30 mbar (Figure~\ref{temp2}) for four models---GCM5, GCM6, GCM7,
and GCM8---initialized with the jet profiles shown in
Figure~\ref{gcmjet}.  Except for the initial conditions, everything
about the models are identical.  Despite the vastly differing initial
conditions, the models again all converge to the same final state.
The equilibrated state exhibits a fast ($3\rm\,km\,s^{-1}$) equatorial
jet, with day-night temperature differences that are hundreds of K at
low pressure.  Interestingly, because of the longer radiative time
constant at the top, the day-night temperature difference in
GCM5--GCM8 is smaller than in models GCM1--GCM4, and there is a
larger eastward displacement of the dayside hot region relative
to the substellar point.  Thus, while it is clear that the response of
our hot-Jupiter models do depend on forcing parameters, they do not
depend on initial conditions.

\begin{figure*}[htb!]
\includegraphics[scale=0.8, angle=0]{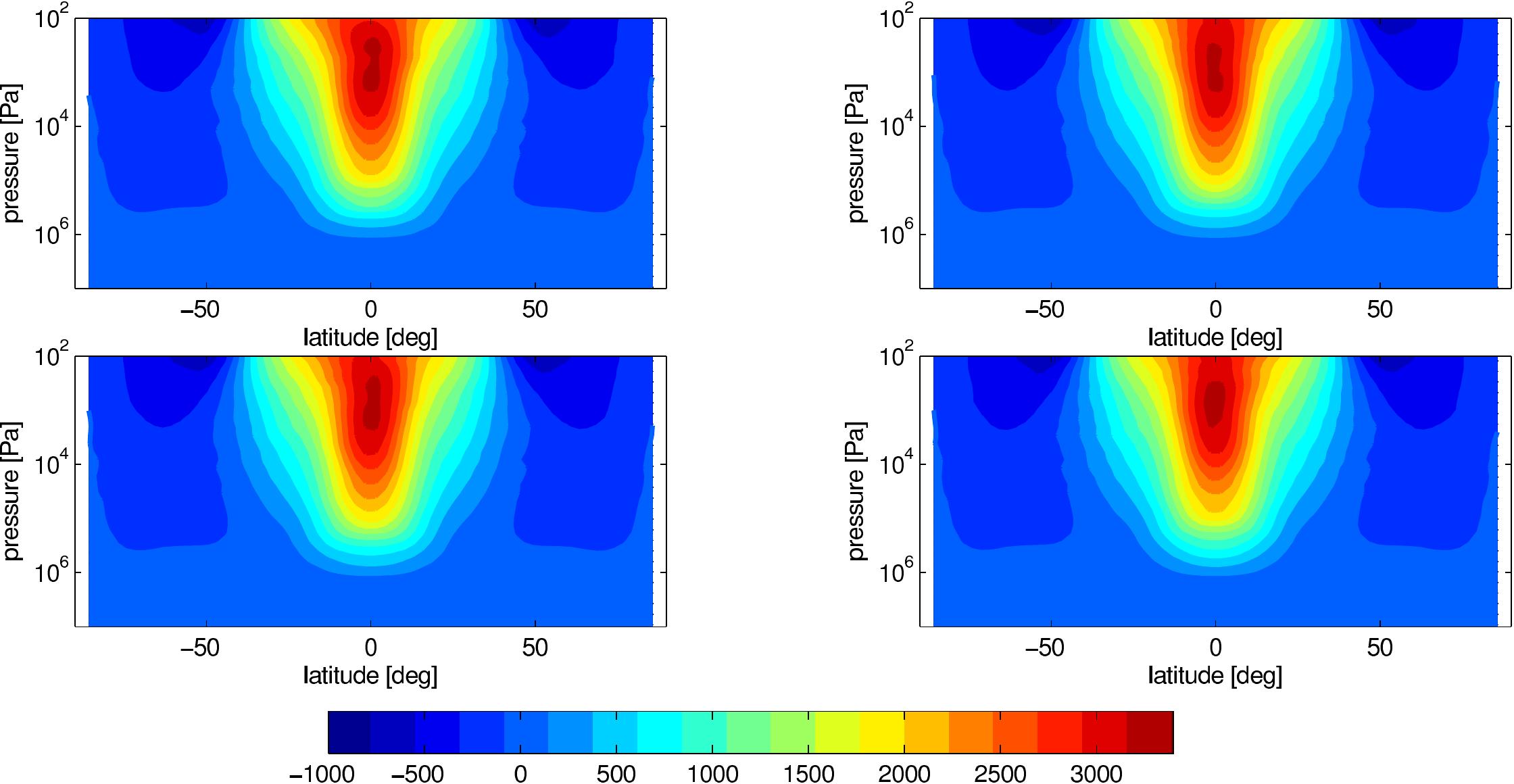}
\caption{ Zonal-mean zonal wind ($\rm m\,s^{-1}$) at four different times for a single
  model, GCM8.  Panels give snapshots at 833 days ({\it top left}),
  903 days ({\it top right}), 972 days ({\it bottom left}), and 1042
  days ({\it bottom right}).  The differences result from time
  variability. These differences are at least as large as the
  model-to-model differences shown in Figure~\ref{u2}, indicating that
  those differences result from time variability rather than
  sensitivity of the statistical steady state to initial conditions.
}
\label{gcm-time}
\end{figure*}

If one compares the panels in Figures~\ref{u2} and \ref{temp2} carefully,
very slight differences become apparent; the peak speed at the core
of the equatorial jet, for example, are not quite identical between
the four panels.  Likewise there are slight, second-order fluctuations in the 
pattern of the velocity vectors in Figure~\ref{temp1}, particularly on
the nightside within $\sim$$40^{\circ}$ latitude of the equator.
These differences are the result of time-variability,
which induces a slight randomness about the statistical steady state,
rather than any fundamental sensitivity to initial conditions.  This
is demonstrated in Figure~\ref{gcm-time}, where the zonal-mean zonal
wind from model GCM8 is shown at four different times after
the model has reached statistical steady state.  The figure shows
that the equatorial jet fluctuates slightly in time; the amplitude
of these fluctuations is comparable to the inter-model differences 
seen in Figure~\ref{u2}.  A temporal average of the temperature
or velocity patterns removes these random fluctuations and yields
a pattern that is essentially identical between models with the
same forcing parameters but differing initial conditions.

We also performed models with weaker damping and drag in the deep
atmosphere, for example models with $\tau_{\rm rad,bot}$ of
$10^7\rm\,s$, $kF$ of $0.01\rm\,day^{-1}$, and/or $p_{\rm drag,top}$
of $10\rm\,bar$ (not shown).  We also performed some models with
weaker day-night forcing, e.g., $\Delta T_{\rm eq}=100\rm\,K$ rather
than 1000 K as for most of the models in this paper.  Likewise, we
also tried some qualitatively different initial conditions, such as
models containing nonzero zonal flow over only a specified subset of
longitudes.  In each case, for a given set of forcing/damping
parameters, the models equilibrate to a statistical steady state with
no sensitivity to initial conditions.

\begin{figure*}[htb!]
\includegraphics[scale=0.29, angle=0]{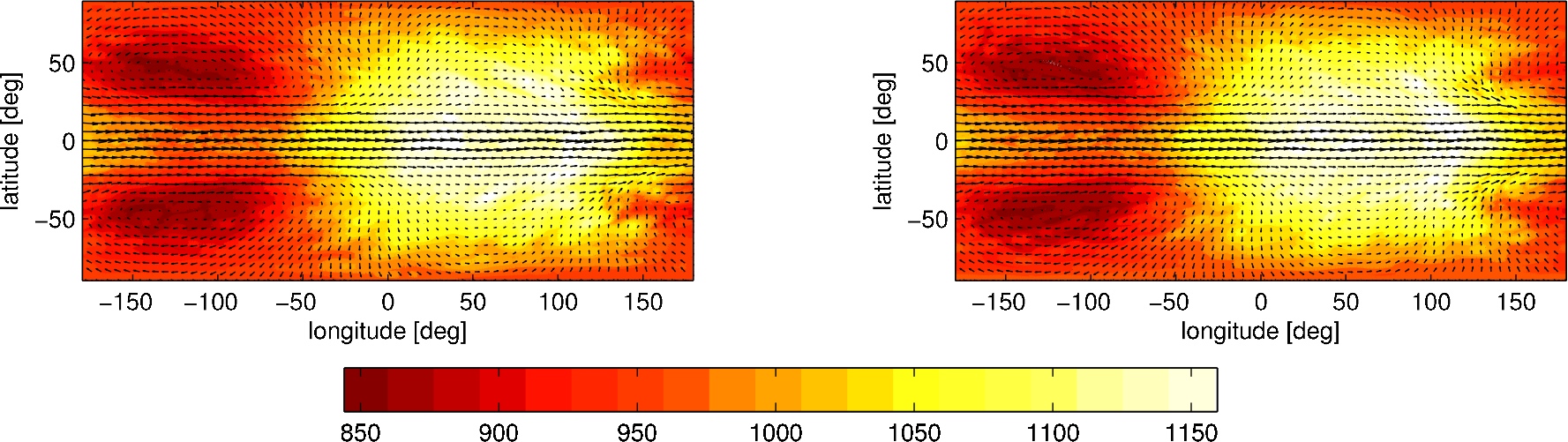}
\vskip 10pt
\includegraphics[scale=0.29, angle=0]{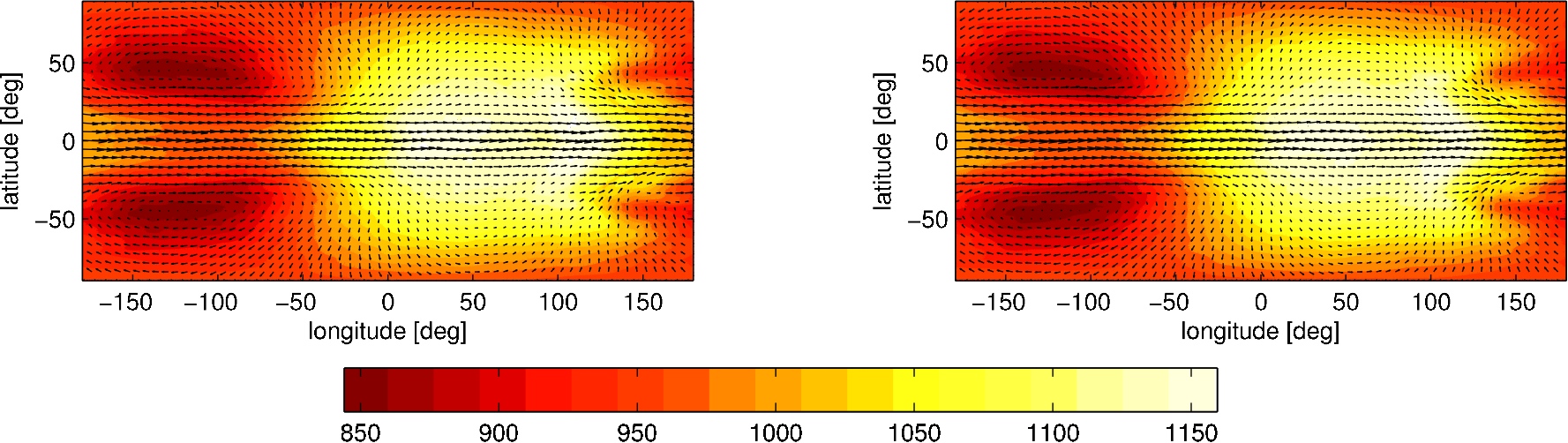}
\vskip 5pt
\includegraphics[scale=0.8, angle=0]{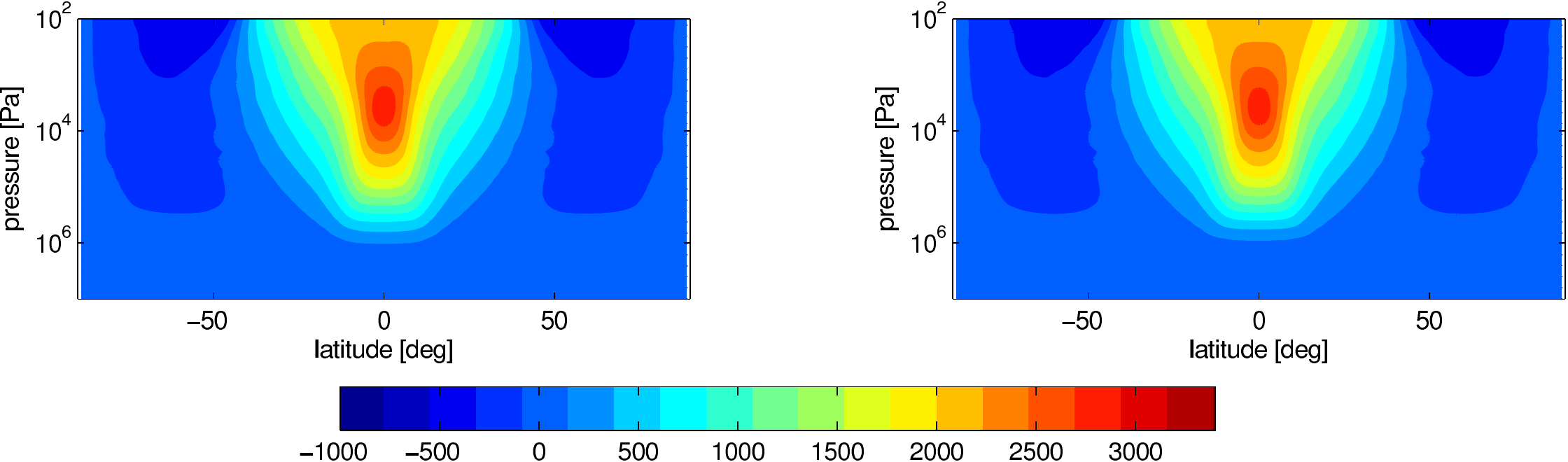}
\caption{Wind and temperature for two otherwise identical
  high-resolution models initialized with differing initial
  conditions.  Resolution is C128, corresponding to a global grid of
  $512\times 256$ in longitude and latitude, with 40 vertical levels.
  Left column shows a model initialized with an initial eastward jet
  (like that in the lower right panel of Figure~\ref{gcmjet}) while
  right column shows a model initialized with an initial westward jet
  (like that in the lower right panel of Figure~\ref{gcmjet} times
  $-1$).  Physical parameters in both models are identical to those in
  low-resolution models GCM5--8 (i.e., $\tau_{\rm
    rad,top}=10^5\rm\,s$, $\tau_{\rm rad,bot}=10^6\rm\,s$,
  $kF=0.1\rm\,day^{-1}$, and $p_{\rm drag,top}=1\rm\,bar$).  {\it Top
    row:} Snapshots of temperature (colorscale, K) and winds (arrows)
  on the 30-mbar level at a time of 1042 days.  {\it Middle row:} Time
  average from 868 to 1042 days of temperature and winds at 30 mbars.
  {\it Bottom row:} Time average from 868 to 1042 days of zonal-mean
  zonal wind ($\rm m\,s^{-1}$).  Models converge to the same
  statistical steady state regardless of initial condition.  The
  statistical steady state is also extremely similar to those obtained
  in low-resolution C32 models with the same parameters (compare to
  Figures~\ref{u2} and \ref{temp2}). }
\label{high-res}
\end{figure*}

The lack of sensitivity to initial conditions in our models is not an
artifact of our model resolution or numerical damping (i.e., the
Shapiro filter) but rather is a fundamental property of the system
behavior over the range explored.  We also integrated models with
resolutions of C64 and C128, approximately equivalent to global
resolutions of $256\times128$ and $512\times256$, respectively, in
longitude and latitude.  To the best of our knowledge, our C128
models, in particular, are higher resolution than any published
three-dimensional models of hot Jupiters that include day-night
thermal forcing to date.  Initial conditions corresponding to rest
states, eastward barotropic jets, and westward barotropic jets
(identical to that shown in the lower right panel of
Figure~\ref{gcmjet} but multiplied by $-1$) were explored.  {\tt
  Figure~\ref{high-res} shows these results for the C128 models
  initialized with an eastward jet (left column) and a westward jet
  (right column).}  All of these models converged to a statistical
steady state that is essentially identical, showing that memory of the
initial condition has been lost.  {\tt Instantaneous snapshots of the
  temperature field are extremely similar in overall structure
  (Figure~\ref{high-res}, top row).  The flow does 
  exhibit some small-scale structure that is time-variable and differs
  from one snapshot to another (either between different simulations
  or at different times of a given simulation).  Averaging in time to
  determine the statistical steady state leads to temperature and wind
  fields that are essentially identical regardless of the initial
  condition (Figure~\ref{high-res}, middle and bottom rows).}
For a given set of forcing parameters, the overall
pattern of temperature and winds are also extremely similar between
our C32, C64, and C128 models, suggesting that numerical convergence
has nearly been reached even by C32 (compare Figure~\ref{high-res}
with Figures~\ref{u2} and \ref{temp2}).

It is interesting to consider the situation where frictional drag is
excluded (i.e., $kF=0$).  Because no mass can enter or leave the model
domain, and the top and bottom boundaries are free-slip in horizontal
momentum, the absence of frictional drag implies the absence of any
external torques acting on the system.  In such a situation, the
globally integrated angular momentum over the domain is conserved to
within numerical accuracy.  Therefore, a drag-free model initialized
with an eastward jet will exhibit a different total angular
momentum---for all time---than a drag-free model initialized from rest
or from a westward jet.  As a result, drag-free models initialized
with differing angular momentum cannot converge to the same final
state, because there is no mechanism to force their differing angular
momenta to converge to a single value.  However, it is important to
emphasize that this situation is artificial: on a real planet the
atmosphere will interact with the interior, leading to a torque on the
atmosphere that allows the atmospheric state to adjust its angular
momentum relative to the interior, and this will remove this
initial-condition sensitivity on the atmospheric flow.  In our
shallow-water models (Section~\ref{2D model}), the active layer
interacts with an underlying (assumed quiescent) interior, and this
explains why these models exhibit no sensitivity to initial condition
even in the case where drag is excluded from the active layer (i.e.,
$\tau_{\rm drag}\to \infty$).  In our 3D models, the application of
frictional drag (i.e., non-zero $kF$) near the bottom is crudely
intended to represent such an atmosphere-interior interaction and
again explains the lack of sensitivity to initial conditions in those
models.  Even in the absence of drag, the existence of a deep,
quiescent, inert layer at the bottom of the domain, as exists in many
published hot-Jupiter models in the literature, can serve as a 
reservoir of mass and momentum that plays a similar role.

\begin{figure*}[htb!]
\includegraphics[scale=0.8, angle=0]{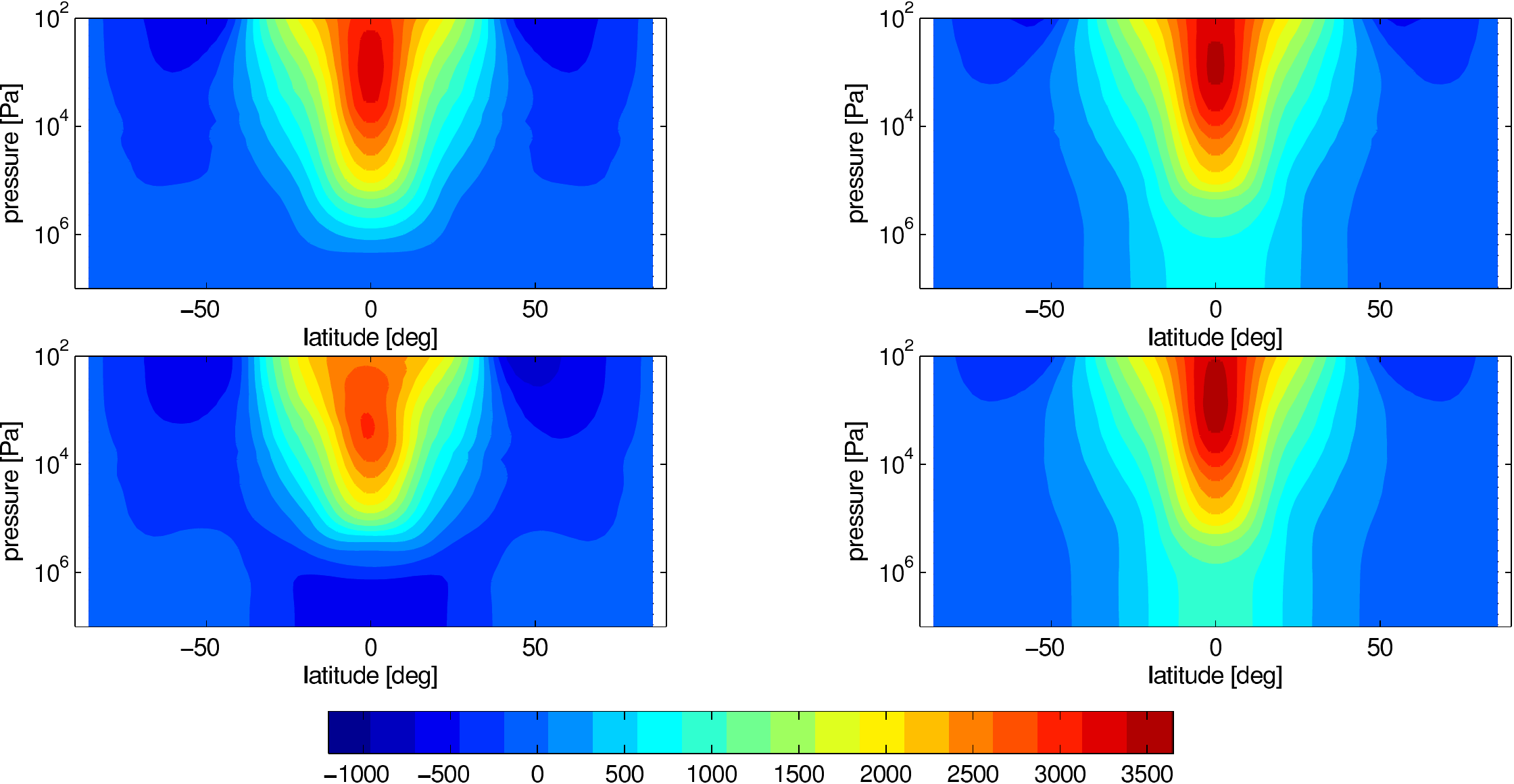}
\caption{Zonal-mean zonal wind ($\rm \,m\,s^{-1}$) in four models with
  no large-scale drag (i.e., $kF=0$) initialized from the four
  different initial conditions given in the corresponding panels of
  Figure~\ref{gcmjet} (rest state in upper left, decaying eastward jet
  in upper right, decaying westward jet in lower left, and barotropic
  eastward jet in lower right).  All other physical and numerical
  parameters are identical to those of GCM5--GCM8.  Because of the
  absence of drag, the angular momenta of the four runs are each
  constant in time (to within numerical errors) but are different from
  each other, and therefore the models do not converge to the same
  steady state.  In particular, memory of the initial jet is retained
  at pressures exceeding 10 bars.  The zonal averages shown here are
  time averaged from 694 to 1042 days.  }
\label{nodrag-ubar}
\end{figure*}

\begin{figure*}[htb!]
\includegraphics[scale=0.8, angle=0]{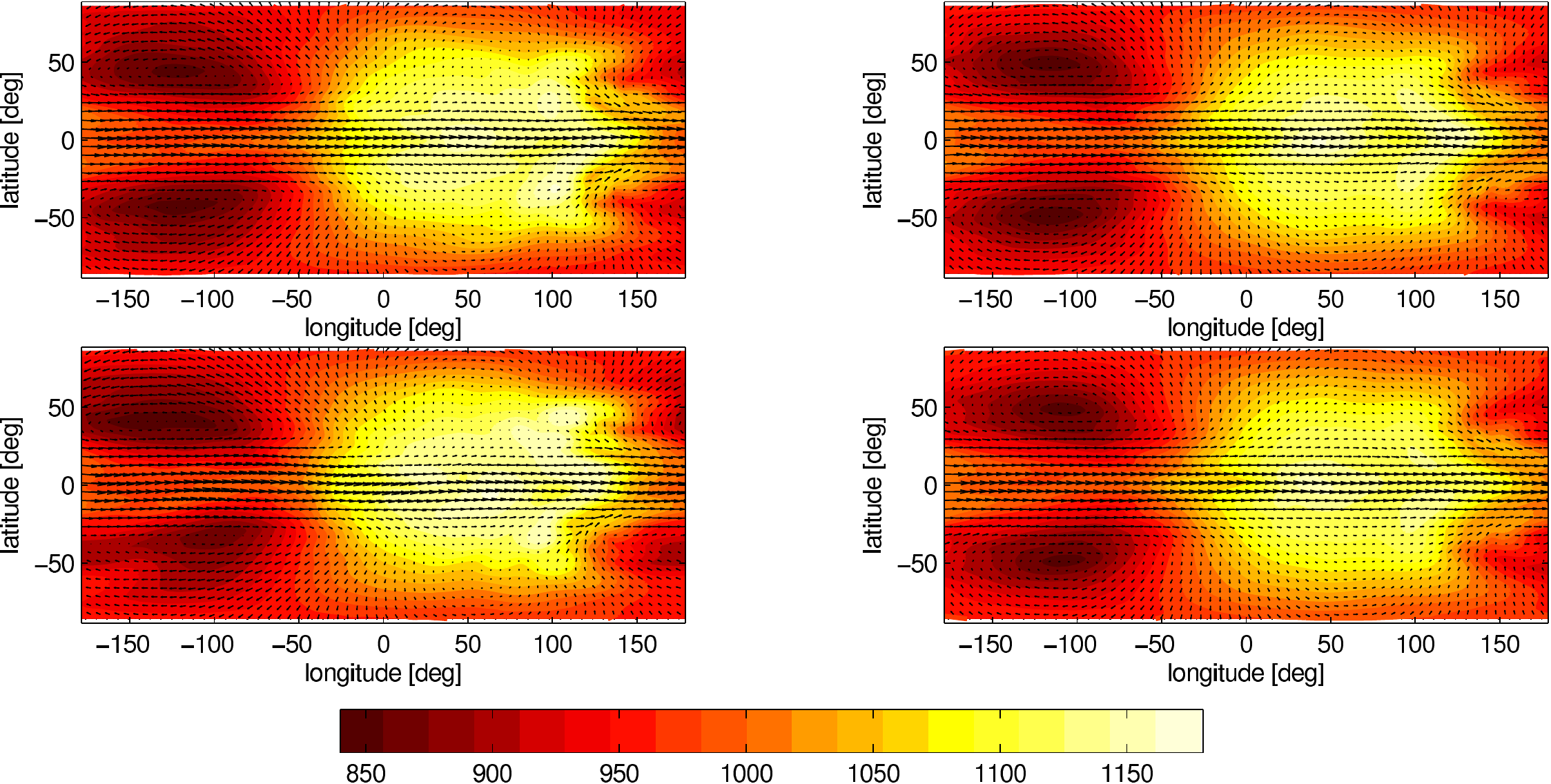}
\caption{Temperature at 30 mbar (color scale, in K) and winds (arrows)
  for the same four drag-free models shown in
  Figure~\ref{nodrag-ubar}.  (The initial conditions of the four
  models correspond to a rest state in upper left, decaying eastward
  jet in upper right, decaying westward jet in lower left, and
  barotropic eastward jet in lower right.)  These are snapshots at
  1042 days.  Interestingly, despite the differences in zonal-jet
  structure at pressures exceeding 10 bars in the four models, the
  temperature structure at 30 mbar---near the infared photosphere on a
  typical hot Jupiter---are quite similar.}
\label{nodrag-temp}
\end{figure*}

Figures~\ref{nodrag-ubar} and \ref{nodrag-temp} illustrate the
behavior when large-scale frictional drag is excluded.  These models
are identical to GCM5--GCM8 in all respects except that the drag
coefficient $kF$ is set to zero (the models still include the Shapiro
filter for numerical stability).  As expected from the arguments
above, the absence of drag means there is no mechanism for the angular
momenta of the four models---which are initially different---to
converge to the same value.  Consistent with this expectation,
Figure~\ref{nodrag-ubar} shows that these models retain memory of the
initial jet at pressures exceeding 10 bars.  Interestingly, however,
the jet profiles at pressures less than 1 bar are quite similar
despite the differing initial conditions.  In the observable
atmosphere, the temperature patterns are likewise very similar for the
different models; this is illustrated in Figure~\ref{nodrag-temp} at
the 30-mbar level, which is near the infrared photosphere for a
typical hot Jupiter.  Because light curve and spectral observations
are determined by the temperature structure at pressures less than 1
bar, this suggests that, in practice, observational predictions are
not strongly sensitive to initial conditions even in this drag-free
case, at least for the range of initial conditions considered here.  We reiterate,
however, that atmosphere-interior interaction on a real hot Jupiter
would be expected to eliminate this sensitivity, as shown in
our shallow-water models and in our 3D models with non-zero $kF$.

We also explored models where the radiative equilibrium temperature
and radiative time constant are independent of depth, as in
\citet{thrastarson-cho-2010}.  These models exhibit significant
large-amplitude time variability that is qualitatively different from
the other models presented in this paper.  When large-scale drag is
included at the bottom of the domain (i.e., non-zero $kF$), time
averages of these solutions show that the statistical steady states
are essentially identical regardless of the initial condition
employed---despite the strong time variability.  When drag is
excluded, we find, as described above, that models whose initial
conditions exhibit different angular momentum are unable to converge
to the same time-mean state.  Regardless, interaction between the flow
and the bottom boundary seems to play a crucial role in the dynamics
when $T_{\rm eq}$ and $\tau_{\rm rad}$ are independent of
pressure---an aspect which is unrealistic for hot Jupiters, whose
atmospheres are not underlain by impermeable surfaces.  When the
radiative equilibrium temperature profiles and radiative time constant
are independent of pressure, the flow exhibits strong horizontal
variations in entropy on the lower boundary.  As is well known, the
existence of horizontal entropy variations against an impermeable
surface tend to make a flow much more baroclinically unstable
\citep[see, e.g.,][Chapter 6]{vallis-2006}.  Such instabilities can
lead to significant time variability, particularly when the Rossby
deformation radius is global in scale, and this may help to explain
the large degree of temporal variability in these models as well as
the models of \citet{thrastarson-cho-2010}.  By comparison, our
nominal models (e.g., GCM1 through GCM8) are set up so that the
thermal forcing and horizontal entropy gradients are weak at the lower
boundary; this helps to avoid such lower-boundary instabilities, which
are artificial in the context of a gas giant.

\section{Discussion and Conclusions}
\label{conclusion}

We explored the sensitivity to initial conditions of three-dimensional
models of synchronously rotating hot Jupiters with day-night thermal
forcing.  The thermal forcing was chosen to be strong at the top and
weak at the bottom, as must occur on real hot Jupiters.  Models were
integrated from rest and from various eastward and westward jet
profiles with speeds up to $\sim$$1.5\rm\,km\,s^{-1}$.  In all models
explored, we found that the statistical steady states are independent
of initial conditions---as long as the flow is anchored by interaction
with a planetary interior so that the angular momentum of the
atmosphere can adjust relative to that of the interior.  In the
context of atmosphere models, such interaction could be parameterized
by frictional drag near the bottom of the domain or by allowing the
atmosphere to exchange mass, energy, and angular momentum with a
specified abyssal layer underlying the atmosphere.  When such an
interaction is included, all our models---for a given set of forcing
parameters---converged to the same statistical
steady state regardless of the initial condition employed.  When the
thermal forcing is strong, the circulation in the equilibrated state
exhibits modest time variability that induces small-amplitude random
fluctuations in any given realization.  The statistical steady state
itself, including not only the time-mean wind and temperature but the
overall amplitude of these fluctuations, are independent of initial
conditions when drag---or direct interaction with a specified abyssal
layer---are included.

As described in the Section~\ref{Introduction}, the issue of
initial-condition sensitivity is perhaps best thought of in terms of
whether the atmospheric circulation exhibits a single, rather than
multiple, stable equilibria.  Taken at face value, our models 
suggest empirically that, for any given set of forcing and damping parameters,
there exists only one stable equilibrium---at least for the range of
forcing, damping, and planetary parameters explored here.  We have
intentionally chosen forcing and planetary parameters similar to those
appropriate to typical hot Jupiters, including HD 189733b and HD
209458b, as explored by a number of authors \citep{showman-etal-2009,
heng-etal-2011, heng-etal-2011b, perna-etal-2012, rauscher-menou-2012}.
Therefore, we expect our fundamental result---the lack of sensitivity
to initial conditions---to apply generally to the regimes explored
in those papers. 

In this context, it is interesting that our findings differ so
drastically from those of \citet{thrastarson-cho-2010}.  Their models
differ from ours in two important ways: they lack large-scale drag,
and the profiles of $T_{\rm eq}$ and $\tau_{\rm rad}$ in their
Newtonian-cooling scheme are independent of pressure.  Both these
differences seem to contribute to the differences in their results
relative to those presented here.  In particular, the absence of
frictional drag in a 3D model with free-slip boundary conditions and
no mass fluxes through the boundaries implies the absence of any
external torques that could change the globally integrated angular
momentum over time.  Thus, the globally integrated angular momentum of
such a model will be conserved over time to within numerical accuracy.
Since initial conditions corresponding to eastward jets, westward
jets, and rest states exhibit different angular momenta, there is thus
no mechanism to force those models to converge to the same angular
momentum and hence final state.  In practice, we found that this
sensitivity is not strong in the observable atmosphere for the range
of initial conditions explored here.  Nevertheless, it may be stronger
when $T_{\rm eq}$ and $\tau_{\rm rad}$ are constant with depth, as is
the case in most of \citet{thrastarson-cho-2010}'s models.

Regardless, our results highlight the importance of anchoring the flow
to an assumed planetary interior, either via the application of
frictional drag; the introduction of a deep, quiescent layer at the
bottom of the domain (as exists in many published 3D hot Jupiter
models), or the explicit assumption of an abyssal layer underlying the
active layer (as exists in 1-1/2 layer shallow-water models).  Only in
this way can the atmosphere experience net torques that allow it to
adjust angular momentum over time, allowing models with differing
initial angular momenta to converge to a single statistical steady
state.


\vskip 5pt

Overall, our results indicate that specification of initial conditions
is not a source of uncertainty in atmospheric circulation models of
hot Jupiters, at least over the parameter range explored here.  This
supports the continued use of hot-Jupiter GCMs for understanding
dynamical mechanisms, explaining observations, and making predictions
to help guide future observations.  That said, our results, as well as
those of numerous previous publications, show that details of the
radiative forcing and frictional damping significantly affect the
flow structure, including the qualitative dynamical regime,
wind speeds, day-night temperature differences, and longitudinal
offsets of any hot or cold regions.    State-of-the-art GCMs now
exist that include detailed non-grey radiative transfer
\citep{showman-etal-2009} as well as simpler, faster, gray treatments
\citep{heng-etal-2011b, rauscher-menou-2012, perna-etal-2012}.  By
comparison, our understanding of how to specify frictional drag
is less well developed, and areas such as inclusion of clouds,
sub-gridscale parameterizations of turbulent mixing, and coupling
to chemistry have received little attention.  Continued model development
in these areas, and comparison of such models to observations,
should improve our ability to discern the physical and dynamical
regimes of these fascinating worlds.


\vskip 20pt

\acknowledgements
We thank the International Summer Institute for Modeling
in Astrophysics (ISIMA), and its organizers, particularly Pascale
Garaud and Doug Lin, for a stimulating environment in which to begin
this project.  We also thank Yuan Lian, Lorenzo Polvani, and Ray
Pierrehumbert for useful discussions.

\bibliographystyle{apj}
\bibliography{showman-bib}

\begin{thebibliography}{46}
\expandafter\ifx\csname natexlab\endcsname\relax\def\natexlab#1{#1}\fi

\bibitem[{{Adcroft} {et~al.}(2004){Adcroft}, {Campin}, {Hill}, \&
  {Marshall}}]{adcroft-etal-2004}
{Adcroft}, A., {Campin}, J.-M., {Hill}, C., \& {Marshall}, J. 2004, Monthly
  Weather Review, 132, 2845

\bibitem[{{Arakawa} \& {Lamb}(1977)}]{arakawa-lamb-1977}
{Arakawa}, A., \& {Lamb}, V. 1977, Methods in Computational Physics, 17, 173

\bibitem[{{Budyko}(1969)}]{budyko-1969}
{Budyko}, M.~I. 1969, Tellus, 21, 611

\bibitem[{{Charbonneau} {et~al.}(2008){Charbonneau}, {Knutson}, {Barman},
  {Allen}, {Mayor}, {Megeath}, {Queloz}, \& {Udry}}]{charbonneau-etal-2008}
{Charbonneau}, D., {Knutson}, H.~A., {Barman}, T., {Allen}, L.~E., {Mayor}, M.,
  {Megeath}, S.~T., {Queloz}, D., \& {Udry}, S. 2008, \apj, 686, 1341

\bibitem[{{Cho} \& {Polvani}(1996)}]{cho-polvani-1996a}
{Cho}, J.~Y.-K., \& {Polvani}, L.~M. 1996, Science, 8, 1

\bibitem[{{Cooper} \& {Showman}(2005)}]{cooper-showman-2005}
{Cooper}, C.~S., \& {Showman}, A.~P. 2005, \apjl, 629, L45

\bibitem[{{Cooper} \& {Showman}(2006)}]{cooper-showman-2006}
---. 2006, \apj, 649, 1048

\bibitem[{{Cowan} {et~al.}(2007){Cowan}, {Agol}, \&
  {Charbonneau}}]{cowan-etal-2007}
{Cowan}, N.~B., {Agol}, E., \& {Charbonneau}, D. 2007, \mnras, 379, 641

\bibitem[{{Dobbs-Dixon} {et~al.}(2010){Dobbs-Dixon}, {Cumming}, \&
  {Lin}}]{dobbs-dixon-etal-2010}
{Dobbs-Dixon}, I., {Cumming}, A., \& {Lin}, D.~N.~C. 2010, \apj, 710, 1395

\bibitem[{{Dobbs-Dixon} \& {Lin}(2008)}]{dobbs-dixon-lin-2008}
{Dobbs-Dixon}, I., \& {Lin}, D.~N.~C. 2008, \apj, 673, 513

\bibitem[{{Dowling} \& {Ingersoll}(1989)}]{dowling-ingersoll-1989}
{Dowling}, T.~E., \& {Ingersoll}, A.~P. 1989, Journal of Atmospheric Sciences,
  46, 3256

\bibitem[{{Hack} \& {Jakob}(1992)}]{hack-jakob-1992}
{Hack}, J.~J., \& {Jakob}, R. 1992, Description of a global shallow water model
  based on the spectral transform method, Tech. rep., National Center for
  Atmospheric Research Technical note NCAR/TN-343+STR, Boulder, CO

\bibitem[{{Harrington} {et~al.}(2006){Harrington}, {Hansen}, {Luszcz},
  {Seager}, {Deming}, {Menou}, {Cho}, \& {Richardson}}]{harrington-etal-2006}
{Harrington}, J., {Hansen}, B.~M., {Luszcz}, S.~H., {Seager}, S., {Deming}, D.,
  {Menou}, K., {Cho}, J.~Y.-K., \& {Richardson}, L.~J. 2006, Science, 314, 623

\bibitem[{{Harrington} {et~al.}(2007){Harrington}, {Luszcz}, {Seager},
  {Deming}, \& {Richardson}}]{harrington-etal-2007}
{Harrington}, J., {Luszcz}, S., {Seager}, S., {Deming}, D., \& {Richardson},
  L.~J. 2007, \nat, 447, 691

\bibitem[{{Held} \& {Suarez}(1994)}]{held-suarez-1994}
{Held}, I.~M., \& {Suarez}, M.~J. 1994, Bulletin of the American Meteorological
  Society, vol.~75, Issue 10, pp.1825-1830, 75, 1825

\bibitem[{{Heng} {et~al.}(2011{\natexlab{a}}){Heng}, {Frierson}, \&
  {Phillipps}}]{heng-etal-2011b}
{Heng}, K., {Frierson}, D.~M.~W., \& {Phillipps}, P.~J. 2011{\natexlab{a}},
  \mnras, 418, 2669

\bibitem[{{Heng} {et~al.}(2011{\natexlab{b}}){Heng}, {Menou}, \&
  {Phillipps}}]{heng-etal-2011}
{Heng}, K., {Menou}, K., \& {Phillipps}, P.~J. 2011{\natexlab{b}}, \mnras, 413,
  2380

\bibitem[{{Iro} {et~al.}(2005){Iro}, {B{\'e}zard}, \&
  {Guillot}}]{iro-etal-2005}
{Iro}, N., {B{\'e}zard}, B., \& {Guillot}, T. 2005, \aap, 436, 719

\bibitem[{{Knutson} {et~al.}(2008){Knutson}, {Charbonneau}, {Allen}, {Burrows},
  \& {Megeath}}]{knutson-etal-2008}
{Knutson}, H.~A., {Charbonneau}, D., {Allen}, L.~E., {Burrows}, A., \&
  {Megeath}, S.~T. 2008, \apj, 673, 526

\bibitem[{{Knutson} {et~al.}(2007){Knutson}, {Charbonneau}, {Allen}, {Fortney},
  {Agol}, {Cowan}, {Showman}, {Cooper}, \& {Megeath}}]{knutson-etal-2007b}
{Knutson}, H.~A., {et~al.} 2007, \nat, 447, 183

\bibitem[{{Lewis} {et~al.}(2010){Lewis}, {Showman}, {Fortney}, {Marley},
  {Freedman}, \& {Lodders}}]{lewis-etal-2010}
{Lewis}, N.~K., {Showman}, A.~P., {Fortney}, J.~J., {Marley}, M.~S.,
  {Freedman}, R.~S., \& {Lodders}, K. 2010, \apj, 720, 344

\bibitem[{{Li} \& {Goodman}(2010)}]{li-goodman-2010}
{Li}, J., \& {Goodman}, J. 2010, ArXiv e-prints

\bibitem[{{Marley} \& {McKay}(1999)}]{marley-mckay-1999}
{Marley}, M.~S., \& {McKay}, C.~P. 1999, Icarus, 138, 268

\bibitem[{{Mayor} \& {Queloz}(1995)}]{mayor-queloz-1995}
{Mayor}, M., \& {Queloz}, D. 1995, \nat, 378, 355

\bibitem[{{Menou}(2012)}]{menou-2012}
{Menou}, K. 2012, \apj, 745, 138

\bibitem[{{Menou} \& {Rauscher}(2009)}]{menou-rauscher-2009}
{Menou}, K., \& {Rauscher}, E. 2009, \apj, 700, 887

\bibitem[{{Perna} {et~al.}(2012){Perna}, {Heng}, \& {Pont}}]{perna-etal-2012}
{Perna}, R., {Heng}, K., \& {Pont}, F. 2012, \apj, 751, 59

\bibitem[{{Perna} {et~al.}(2010){Perna}, {Menou}, \&
  {Rauscher}}]{perna-etal-2010}
{Perna}, R., {Menou}, K., \& {Rauscher}, E. 2010, \apj, 719, 1421

\bibitem[{{Pierrehumbert}(2010)}]{pierrehumbert-2010}
{Pierrehumbert}, R.~T. 2010, {Principles of Planetary Climate} (Cambridge
  University Press)

\bibitem[{{Polvani} {et~al.}(1995){Polvani}, {Waugh}, \&
  {Plumb}}]{polvani-etal-1995}
{Polvani}, L.~M., {Waugh}, D.~W., \& {Plumb}, R.~A. 1995, Journal of
  Atmospheric Sciences, 52, 1288

\bibitem[{{Rauscher} \& {Menou}(2010)}]{rauscher-menou-2010}
{Rauscher}, E., \& {Menou}, K. 2010, \apj, 714, 1334

\bibitem[{{Rauscher} \& {Menou}(2012)}]{rauscher-menou-2012}
---. 2012, \apj, 745, 78

\bibitem[{{Scott} \& {Polvani}(2007)}]{scott-polvani-2007}
{Scott}, R.~K., \& {Polvani}, L. 2007, J. Atmos. Sci, 64, 3158

\bibitem[{{Scott} \& {Polvani}(2008)}]{scott-polvani-2008}
{Scott}, R.~K., \& {Polvani}, L.~M. 2008, \grl, 35, L24202

\bibitem[{{Showman}(2007)}]{showman-2007}
{Showman}, A.~P. 2007, J. Atmos. Sci., 64, 3132

\bibitem[{{Showman} {et~al.}(2008){Showman}, {Cooper}, {Fortney}, \&
  {Marley}}]{showman-etal-2008a}
{Showman}, A.~P., {Cooper}, C.~S., {Fortney}, J.~J., \& {Marley}, M.~S. 2008,
  \apj, 682, 559

\bibitem[{{Showman} {et~al.}(2012){Showman}, {Fortney}, \&
  {Lewis}}]{showman-etal-2012}
{Showman}, A.~P., {Fortney}, J.~J., \& {Lewis}, N. 2012, \apj, submitted to ApJ

\bibitem[{{Showman} {et~al.}(2009){Showman}, {Fortney}, {Lian}, {Marley},
  {Freedman}, {Knutson}, \& {Charbonneau}}]{showman-etal-2009}
{Showman}, A.~P., {Fortney}, J.~J., {Lian}, Y., {Marley}, M.~S., {Freedman},
  R.~S., {Knutson}, H.~A., \& {Charbonneau}, D. 2009, \apj, 699, 564

\bibitem[{{Showman} \& {Guillot}(2002)}]{showman-guillot-2002}
{Showman}, A.~P., \& {Guillot}, T. 2002, \aap, 385, 166

\bibitem[{{Showman} \& {Polvani}(2010)}]{showman-polvani-2010}
{Showman}, A.~P., \& {Polvani}, L.~M. 2010, \grl, 37, 18811

\bibitem[{{Showman} \& {Polvani}(2011)}]{showman-polvani-2011}
---. 2011, \apj, 738, 71

\bibitem[{{Thrastarson} \& {Cho}(2010)}]{thrastarson-cho-2010}
{Thrastarson}, H.~T., \& {Cho}, J. 2010, \apj, 716, 144

\bibitem[{{Thrastarson} \& {Cho}(2011)}]{thrastarson-cho-2011}
{Thrastarson}, H.~T., \& {Cho}, J.~Y. 2011, \apj, 729, 117

\bibitem[{{Vallis}(2006)}]{vallis-2006}
{Vallis}, G.~K. 2006, Atmospheric and Oceanic Fluid Dynamics: Fundamentals and
  Large-Scale Circulation (Cambridge Univ. Press, Cambridge, UK)

\bibitem[{{Watkins} \& {Cho}(2010)}]{watkins-cho-2010}
{Watkins}, C., \& {Cho}, J. 2010, \apj

\bibitem[{{Wright} {et~al.}(2011){Wright}, {Fakhouri}, {Marcy}, {Han}, {Feng},
  {Johnson}, {Howard}, {Fischer}, {Valenti}, {Anderson}, \&
  {Piskunov}}]{wright-etal-2011}
{Wright}, J.~T., {et~al.} 2011, \pasp, 123, 412

\end{thebibliography}

\end{document}